\documentclass[12pt, draftclsnofoot, onecolumn]{IEEEtran}

\usepackage{bigints}
\usepackage{color}  
\usepackage[dvipsnames]{xcolor}
\usepackage[final]{graphicx} 
\usepackage{amsmath,amssymb,amscd,latexsym,dsfont,amsthm,algorithm,,amsbsy,epsfig,bbm,mathrsfs,fancyhdr,fancyvrb}
\usepackage{upgreek,textgreek}
\usepackage{subcaption}
\usepackage{caption}
\usepackage{psfrag}
\usepackage{epstopdf} 
\usepackage{tabls}
\usepackage{cite}
\usepackage{color,comment}
\usepackage{balance}
\usepackage{float}
\usepackage{placeins}
\usepackage{lipsum}
\usepackage{algorithm,algpseudocode}
\usepackage{algorithmicx}
\usepackage{tikz}
\usepackage[percent]{overpic}
\usetikzlibrary{calc}
\usepackage{hyperref}
\usepackage{relsize}
\usepackage{array}
\usepackage{booktabs}

\newcommand{\tr}[1]{\mathrm{#1}}

\newcommand\norm[1]{\left\lVert#1\right\rVert}

\usetikzlibrary{decorations.pathreplacing}

\tikzset{radiation/.style={{decorate,decoration={expanding waves,angle=90,segment length=2pt}}},
         relay/.pic={
        code={\tikzset{scale=5/10}
            \draw[semithick] (0,0) -- (1,4);
            \draw[semithick] (3,0) -- (2,4);
            \draw[semithick] (0,0) arc (180:0:1.5 and -0.5) node[above, midway]{#1};
            \node[inner sep=4pt] (circ) at (1.5,5.5) {};
            \draw[semithick] (1.5,5.5) circle(8pt);
            \draw[semithick] (1.5,5.5cm-8pt) -- (1.5,4);
            \draw[semithick] (1.5,4) ellipse (0.5 and 0.166);
            \draw[semithick,radiation,decoration={angle=45}] (1.5cm+8pt,5.5) -- +(0:2);
            \draw[semithick,radiation,decoration={angle=45}] (1.5cm-8pt,5.5) -- +(180:2);
  }},
  }
 \tikzset{ pics/.cd,
 	SU/.style={code={
    \begin{scope}[local bounding box=#1]
      \fill [even odd rule, pic actions/.try]
        (-1,-5/2) -- (-1,-1/8) -- (1,-1/8) -- (1,-5/2)
        arc (360:180:1 and 1/4) -- cycle
     (-1,5/2) -- (-1,1/8) -- (1,1/8) -- (1,5/2)
        arc (0:180:1 and 1/4) -- cycle
     (-3/4, 9/4) -- (-3/4, 3/8) -- (3/4, 3/8) -- (3/4, 9/4)
     arc (0:180:3/4 and 1/8)-- cycle
     \foreach \i in {-1,0,1}{\foreach \j in {1,2,3}{
       (-\i*1/2-3/16,-\j/2-3/4) rectangle ++(3/8, 3/8)}}
     (-1/2,-3/4) rectangle (1/2, -1/4);
   \end{scope}
  }},
}

\newcounter{tempEquationCounter}
\newcounter{thisEquationNumber}
{\setcounter{equation}{\value{tempEquationCounter}}
\end{figure*}
}%

\newtheorem{proposition}{Proposition}

\algdef{SE}[DOWHILE]{Do}{doWhile}{\algorithmicdo}[1]{\algorithmicwhile\ #1}%

\begin{document}

			\title{{Interference Management in NOMA-based Fog-Radio Access Networks via Joint Scheduling and Power Adaptation}}

\author{
Itsikiantsoa Randrianantenaina, \textit{Student Member, IEEE}, Megumi Kaneko, \textit{Senior Member, IEEE}, Hayssam Dahrouj, \textit{Senior Member, IEEE}, Hesham ElSawy, \textit{Senior Member, IEEE} and Mohamed-Slim~Alouini, \textit{Fellow, IEEE}\\
\thanks{%
Part of this paper is presented at the IEEE Global Communications Conference (GLOBECOM' 2018), Abu Dhabi, UAE \cite{Tsiky2018}.\newline
\indent I. Randrianantenaina and M.-S. Alouini are with the Computer, Electrical, and Mathematical Science and Engineering (CEMSE) Division, King Abdullah University of Science and Technology (KAUST), Thuwal, Saudi Arabia. M. Kaneko is with the National Institute of Informatics (NII), Tokyo, Japan. H. Dahrouj is with the department of Electrical and Computer Engineering at Effat University, Jeddah, Saudi Arabia. H. ElSawy is with King Fahd University of Petroleum and Minerals (KFUPM), Dhahran, Saudi Arabia. [e-mails: \{itsikiantsoa.randrianantenaina, slim.alouini\}@kaust.edu.sa], megkaneko@nii.ac.jp, hayssam.dahrouj@gmail.com, hesham.elsawy@kfupm.edu.sa}%
\thanks{
}
}
				\maketitle
			\thispagestyle{empty}

\vspace{-1.5cm}
	\begin{abstract}	
Non-Orthogonal Multiple Access (NOMA) and Fog Radio Access Networks (FRAN) are promising candidates within the 5G and beyond systems. This work examines the benefit of adopting NOMA in an FRAN architecture with constrained capacity fronthaul. The paper proposes methods for optimizing joint scheduling and power adaptation in the downlink of a NOMA-based FRAN with multiple resource blocks (RB). We consider a mixed-integer optimization problem which maximizes a network-wide rate-based utility function subject to fronthaul-capacity constraints, so as to determine i) the user-to-RB assignment, ii) the allocated power to each RB, and iii) the power split levels of the NOMA users in each RB. The paper proposes a feasible decoupled solution for such non-convex optimization problem using a three-step hybrid centralized/distributed approach. The proposed solution complies with FRAN operation that aims to partially shift the network control to the FAPs, so as to overcome delays due to fronthaul rate constraints. The paper proposes and compares two distinct methods for solving the assignment problem, namely the Hungarian method, and the Multiple Choice Knapsack method. The power allocation and the NOMA power split optimization, on the other hand, are solved using the alternating direction method of multipliers (ADMM). Simulations results illustrate the advantages of the proposed methods compared to different baseline schemes including the conventional Orthogonal Multiple Access (OMA), for different utility functions and different network environments.
	\end{abstract}
	
	
\section{Overview}
			Due to the exponential growth in number of users and applications that need to be served by next generation wireless systems (5G and beyond), developing efficient utilization schemes of the available radio resources becomes imperative. In conventional cellular networks, each basic radio resource unit - that is time, frequency, code or space - is allocated to a unique user, a strategy that is generically referred to as Orthogonal Multiple Access (OMA). The suboptimality of such an orthogonal allocation, however, has been well established in the information-theoretical results of~\cite{Cover2006}, as larger rate regions can be achieved by simultaneously serving multiple users over each resource unit.
Such principle, also known as Superposition Coding (SC) with Successive Interference Cancelation (SIC), forms the platform of Non-Orthogonal Multiple Access (NOMA), and is known to be capacity-achieving scheme in the Gaussian broadcast channel~\cite{Cover2006}. In NOMA, distinct users messages are superposed in one basic resource unit, and multiplexed in the power domain both by exploiting the channel gain difference between users and by applying SIC~\cite{Islam2017}. In {downlink (DL) NOMA}, substantial rate-fairness trade-off can be achieved by an uneven power split \cite{Tse2005}. While {a user with weak channel power gain,  hereafter denoted as weak user, is served with high power, a user with strong channel power gain,  hereafter denoted as strong user,  is served with low power.} On one hand, the strong user performs SIC decoding, i.e., first, it decodes the weak user's signal and subtracts it from the received signal, and then decodes its own signal. On the other hand, the weak user directly decodes its signal from the received signal by treating the strong user's signal (with low power) as noise. Such applicability of NOMA in 5G systems has recently garnered much interest in the literature of single cell~\cite{Ding2014,Gu2018},  relay networks~\cite{Kaneko2014,Kaneko2015} and large-scale networks~\cite{Ali2017}. The results in~\cite{Ali2017} show that, combined with well-designed interference management, NOMA can yield better spectral efficiency compared to  OMA in large-scale 5G cellular networks.
	
To mitigate mutual interference in ultra-dense 5G networks, Cloud-Radio Access Network (CRAN) incorporates cloud computing capabilities into wireless networks. In CRAN, base-station functionalities are split into two components: Remote Radio Heads (RRHs) which serve as the data plane that grants wireless coverage, and a centralized pool of Baseband Units (BBUs) which acts as the control plane that handles large-scale signal processing and resource management~\cite{Peng2015}. Under such realm, RRHs are connected to the BBU pool through capacity-limited fronthaul links. Resource allocation schemes are abundant in the recent literature of CRANs, e.g., the power allocation and beamforming problem in \cite{Dai2015}. Such schemes, however, are often constrained by the coupled operation of the capacity-limited fronthaul links and the centralized processing schemes, which causes significant transport delays and are unsuitable for the envisioned 5G applications. To {mitigate} the traffic burden on the fronthaul link, an alternative network architecture, named as FRAN, also known as Mobile Edge Computing (MEC) or Cloudlets, recently emerges in network edge operation to perform distributed signal processing and resource management, and to temporarily store the processed data~\cite{Peng2016}\cite{Pontois2018}.

This paper considers the downlink of a NOMA-based FRAN, where the cloud is connected to the fog-access points (FAPs) through capacity-limited backhaul links. The data of every FAP are transmitted over multiple Resource Blocks (RBs), where two users are multiplexed within each RB. The paper then tackles the resource allocation problem of jointly optimizing the RBs-to-users assignment, the power allocation and the NOMA power split by maximizing the weighted-sum rate across the network.

\subsection{Related Work}		
	The integration of NOMA in a CRAN architecture is analyzed in few recent works, e.g., \cite{Gu2018,Vega2017,Vien2017,Hao2018, Lee2018, Singh2017,Vien2015,Zhao2018}.
		References~\cite{Gu2018} and \cite{Vega2017} analyze the outage probability in a NOMA-based CRAN network.  
		In particular, reference \cite{Vega2017} proposes an architecture, where the cell-center users are served by their nearest RRHs as their strong users, while multiple RRHs  collaborate to serve the cell-edge users as their weak users. Using stochastic geometric tools, it is shown in \cite{Vega2017} that NOMA enhances the performance of the cell-edge users.
		
		References~\cite{Vien2017} and \cite{Hao2018} analyze the energy efficiency of the DL of a CRAN cellular network. In ~\cite{Vien2017}, NOMA is adopted in the wireless backhaul links connecting the BSs to the cloud, while in~\cite{Hao2018}, it is adopted in the communication between RRHs and users. Both papers show that the NOMA scheme can achieve higher energy efficiency compared to the conventional OMA.

Several researchers investigate resource allocation for NOMA-based CRAN~\cite{Lee2018, Singh2017,Vien2015,Zhao2018}.
While reference~\cite{Lee2018} optimizes the beamforming strategy in order to maximize the minimum delivery rate of files requested by users, reference~\cite{Singh2017} proposes a heuristic algorithm that maximizes the network sum-rate by separately assigning users to RRHs and optimizing the transmission power. The authors of \cite{Vien2015} propose a distance-based power allocation algorithm, as well as an algorithm that determines the optimal number of BSs. In the same context, reference~\cite{Zhao2018}  proposes a low complexity algorithm that assigns each group of users to subchannels, and determines the transmission subchannel for every RRH.
		

%
		
		 As for the integration of NOMA in a FRAN, reference \cite{Zhang2018} addresses the resource allocation for NOMA-based FRAN where both FAPs and user equipment are able to store some data. Moreover, D2D communication is enabled between the users. The authors in \cite{Zhang2018} propose a power allocation optimization strategy, where every user maximizes its utility function under power constraints and interference-based pricing function. The users in \cite{Zhang2018} are assigned to subchannels using a many-to-many two-sided matching-based solution. To the best of our knowledge, however, no joint user assignment and power allocation optimization solution has been discussed or analyzed in the past literature.
		
			
\subsection{Contribution}

This paper investigates the resource allocation problem for a downlink NOMA-based FRAN architecture composed of several FAPs, each transmitting over multiple RBs. In each RB, each FAP can serve two users through NOMA, where the user with better channel is referred to as the a strong user, while the user with worse channel is called the weak user\footnote{In theory, NOMA allows the superposition of $K\geq 2$ users per RB, but with much reduced incremental gains as $ K  $grows\cite{Tse2005}. This is why we focus here on solving for $K=2$, the most fundamental and insightful case which is already intricate.}. The power allocated to each RB is, therefore, split between its two assigned users.
Unlike reference \cite{Zhang2018} and the conference version of the current paper \cite{Tsiky2018} which focus on the single RB case, this paper proposes a resource allocation algorithm that jointly: a) assigns users to FAPs and RBs, b) allocates power to every RB of every FAP, and c) splits the power between the pair of users served under NOMA within each RB.
Moreover, the proposed method aims at maximizing the weighted sum-rate of the network, a utility function widely used in the literature that can be adjusted to different purposes (e.g., maximizing a proportional fairness utility function,
or maximizing the network sum-rate) \cite{Weeraddana2012,Christensen2008}.

To this end, we first formulate a mixed integer optimization problem that maximizes the network utility function by jointly adjusting the user-to-FAP-and-RB assignment, the power allocated to each RB, and the NOMA power split within each RB among the two assigned users. The limited  fronthaul capacity and the power budget of each FAP are taken into account. 
The formulated problem also accounts for the FRAN inherent local operation constraint of limiting each user to be served from a single FAP on a given RB~\cite{Peng2015, Peng2016}.
Nevertheless, each user may be served by different FAPs across different RBs, which provides an additional layer of diversity. 

The proposed algorithm iterates between solving the binary part (assignment) for fixed power allocation, and then solving the continuous part (power allocation) for fixed assignment. For comparison purposes, the paper proposes two distinct algorithms for solving the assignment step, namely the Hungarian-based algorithm and the Knapsack-based algorithm. The Hungarian-based assignment can be performed sequentially or in parallel for each RB  but it does not take the fronthaul capacity constraint into consideration (the fronthaul capacity can be handled in the subsequent power allocation step). The Knapsack-based algorithm, on the other hand, assigns users to every FAP and RB while guaranteeing that every FAP's fronthaul capacity is satisfied at every step.
The continuous optimization steps  (i.e., the RBs' power allocation and the NOMA power split parameters) are solved afterwards using the alternating direction method of multipliers (ADMM)~\cite{Wang2017} for its numerical stability. ADMM is also well adapted to the structure of our problem, and provides a relatively simple computational complexity solution to the continuous part of the optimization problem.


Through extensive computer simulations, we analyze and discuss each step of the proposed algorithms, both in terms of convergence and in terms of incremental gains. The effect of different levels of fronthaul capacity on the network performance is also presented. The results of the paper particularly show that, compared to conventional OMA, the proposed NOMA strategy under FRAN constraints increases user fairness without sacrificing network sum-rate.

\subsection{Paper Organization}
  \vspace{-0.1in}
 The remainder of this paper is organized as follows. First, the network model, notations and rate expressions are presented in Sec.~\ref{sec:SystemModel}. The optimization problem is formulated and analyzed in Sec.~\ref{sec:OptimizationProblem}. Then, both assignment algorithms are given in Sec.~\ref{sec:AssignmentAlgorithm}, followed by the power allocation algorithms in Sec.~\ref{sec:PowerAllocation}. An analysis of the algorithm complexity is provided in Sec.~\ref{sec:ComplexityAnalysis}. Simulation results are presented in Sec.~\ref{sec:SimulationResults}. Finally, Sec.~\ref{sec:Conclusion} concludes the paper.

\section{System Model}\label{sec:SystemModel}
\subsection{Network Model}
			Consider a DL FRAN composed of $ F $ single antenna FAPs, where each FAP is connected to the BBU pool via a capacity-limited fronthaul link. All FAPs share and universally reuse $R$ orthogonal RBs and serve $U$ users. Each RB represents the basic resource allocation unit which cannot be decomposed further, and hence is allocated to a unique user in conventional OMA systems (such as OFDMA in  LTE-Advanced). By contrast, in the considered NOMA-based system in this paper, a pair of users - one strong and one weak user - can be served on the same RB using power multiplexing.  Note that both users are subject to interference from the transmissions from all other FAPs on the same RB.
		
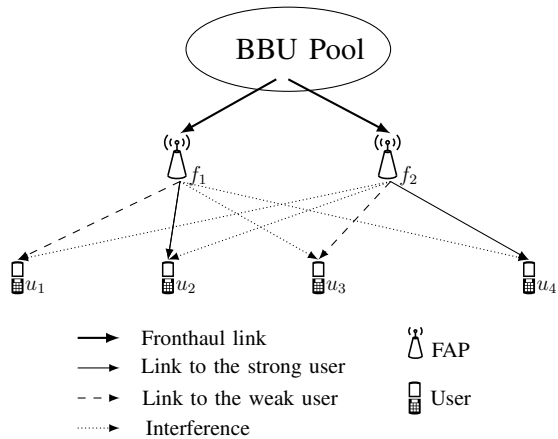
\begin{figure}[t]
				\centering
				\begin{tikzpicture}[scale=0.8, transform shape]
				
				\coordinate (origin) at ($(0,0)$);
				\coordinate (r1) at ($(origin)+(-2,0.9)$);
				\coordinate (r2) at ($(origin)+(1.5,0.9)$);
				\coordinate (r3) at($(origin)+(+3,0)$);
				\coordinate (user1) at ($(origin)+(-4.5,-0.8)$);
				\coordinate (user2) at ($(origin)+(-2,-0.8)$);
				\coordinate (user3) at($(origin)+(0.5,-0.8)$);
				\coordinate (user4) at($(origin)+(4,-0.8)$);
				\coordinate (server1) at($(origin)+(-0.5,3)$);
				\coordinate (server2) at($(origin)+(0.7,3)$);
				\coordinate (legendOrigin) at ($(user1)+(1.5,-1)$);
				
				\node[inner sep=0pt] (ser1) at (server1)
				{};
				\node[inner sep=0pt] (ser2) at (server2)
				{};
				
			\draw[ solid] ($(origin)+(0,3)$) ellipse (1.7cm and 0.7cm);
				
				\path  (r2)  pic[scale=0.2,color=black]{relay= };
				\path  (r1)  pic[scale=0.2,color=black]{relay= };
				
				\pic [ fill=black, scale =0.1]  at (user1) {SU=k};
				\pic [ fill=black, scale =0.1]  at (user2) {SU=k};
				\pic [ fill=black, scale =0.1]  at (user3) {SU=k};
				\pic [ fill=black, scale =0.1]  at (user4) {SU=k};
				
				\draw [-> ,solid, thin,>=latex] ($(r1)+(0.2,-0.1)$) --  ($(user2)+(0,0.3)$) ;
				
				\draw [-> ,solid, thin,>=latex] ($(r1)+(0.2,-0.1)$) --  ($(user2)+(0,0.3)$) ;
				\draw [-> ,solid, thin,>=latex] ($(r2)+(0.2,-0.1)$) --  ($(user4)+(0,0.3)$) ;
				
				
				\draw [-> ,dashed, thin,>=latex] ($(r1)+(0.2,-0.1)$) --  ($(user1)+(0,0.3)$) ;
				\draw [-> ,dashed, thin,>=latex] ($(r2)+(0.2,-0.1)$) --  ($(user3)+(0,0.3)$) ;
				
				\draw [-> ,solid, thick,>=latex] ($(origin)+(-0.2,2.5)$) --  ($(r1)+(0.2,0.7)$) ;
				\draw [-> ,solid, thick,>=latex] ($(origin)+(0,2.5)$) --  ($(r2)+(0.2,0.7)$) ;

				\draw [-> ,densely dotted, thin,>=latex] ($(r1)+(0.2,-0.1)$) --  ($(user3)+(0,0.3)$) ;
				\draw [-> ,densely dotted, thin,>=latex] ($(r2)+(0.2,-0.1)$) --  ($(user1)+(0,0.3)$) ;
				\draw [-> ,densely dotted, thin,>=latex] ($(r2)+(0.2,-0.1)$) --  ($(user2)+(0,0.3)$) ;
				\draw [-> ,densely dotted, thin,>=latex] ($(r1)+(0.2,-0.1)$) --  ($(user4)+(0,0.3)$) ;

				\node [scale=0.85]at ($(r1)+(0.5,+0.)$)  {$f_{1}$};
				\node [scale=0.85]at ($(r2)+(0.5,+0.)$)  {$f_{2}$};
				\node [scale=1.2]at ($(ser2) + (-0.5,0) $)  {BBU Pool};
				\node [scale=0.85]at ($(user1)+(0.3,-0.15)$)  {$u_{1}$};
				\node [scale=0.85]at ($(user2)+(0.3,-0.15)$)  {$u_{2}$};
				\node [scale=0.85]at ($(user3)+(0.3,-0.15)$)  {$u_{3}$};
				\node [scale=0.85]at ($(user4)+(0.3,-0.15)$)  {$u_{4}$};

				\draw [-> ,solid, thick,>=latex] ($(legendOrigin) +(-0.5,0)$) --  ($(legendOrigin) +(0.2,0)$) ;
				\draw [-> ,solid, thin,>=latex] ($(legendOrigin) +(-0.5,-0.5)$) --  ($(legendOrigin) +(0.2,-0.50)$) ;
				\draw [-> ,dashed, thin,>=latex] ($(legendOrigin) +(-0.5,-1)$) --  ($(legendOrigin) +(0.2,-1)$) ;
				\draw [-> ,densely dotted, thin,>=latex] ($(legendOrigin) +(-0.5,-1.5)$) --  ($(legendOrigin) +(0.2,-1.5)$) ;
				
				\path  ($(legendOrigin) +(5,-0.3)$)  pic[scale=0.15,color=black]{relay= };
				\path  ($(legendOrigin) +(5.1,-1)$)  pic[scale=0.1,color=black]{SU=k};
				
				\node [scale=0.85]at ($(legendOrigin) +(1.5+0.1,0)$)  {Fronthaul link};
				\node [scale=0.85]at ($(legendOrigin) +(1.95+0.3,-0.5)$)  {Link to the strong user};
				\node [scale=0.85]at ($(legendOrigin) +(1.9+0.3,-1)$)  {Link to the weak user};
				\node [scale=0.85]at ($(legendOrigin) +(1.4+0.1,-1.5)$)  {Interference};
				
				\node [scale=0.85]at ($(legendOrigin) +(5.7,-0.2)$)  {FAP};
				\node [scale=0.85]at ($(legendOrigin) +(5.7,-1)$)  {User};
				\end{tikzpicture}
				\caption{FRAN model for 4 users served by 2 FAPs for a given RB.}	
				\label{fig:SystemModel}
			\end{figure}

Fig.~\ref{fig:SystemModel} shows an example of the network model for 4 users and 2 FAPs for a given RB. Note that the cloud guarantees synchronization across RBs, i.e., the transmission on different RBs do not interfere with each other.

			%

	\subsection{Notation and Rate Expression}		
			Let $\mathcal{F}=\{1,...,F\}$, $\mathcal{R}=\{1,...,R\}$ and $\mathcal{U}=\{1,...,U\}$ be the sets of FAPs, RBs and users, respectively. The compound channel coefficient that includes pathloss, shadowing, and Rayleigh fading component between FAP $f$ and user $u$ on RB $r$ is denoted by $h_{fru}$.
The total power to serve both strong and weak users by FAP $f$ on RB $r$ is $p_{fr}$. Let $a_{fr}$ and $1-a_{fr}$  be the respective NOMA power split ratios for the strong and weak users, with $0 \leq a_{fr}\leq 1$.
			
			 Let $s_{fru}$ and $w_{fru}$ be two binary variables such that $s_{fru}=1$ if user $u$ is FAP $f$'s strong user on $r$, and 0 otherwise; and $w_{fru}=1$ if user $u$ is  FAP $f$'s weak user on $r$, and 0 otherwise.
			 The useful signal power received by user $u$  from FAP $f$ on RB $r$ is given by
			 			\begin{align}\label{eq:usefulSignal}
			\chi_{fru}\!&=s_{fru}a_{fr}p_{fr}h_{fru}+w_{fru}(1-a_{fr})p_{fr}h_{fru}.
			\end{align}
			\normalsize
		Note that, for every FAP and every RB, a user can be served either as the strong or the weak user (but not both), i.e., $s_{fru}+w_{fru}=1,\ \forall (f,r,u) \in \left(\mathcal{F}\times\mathcal{R}\times\mathcal{U}\right)$. Therefore, only one of the two terms in~\eqref{eq:usefulSignal} is non-zero.
		Every user is served by at most one FAP on a given RB. Thus, the signals from the remaining FAPs on RB $ r $ are treated as interference. The power of the total interference affecting the signal received at user $u$ from FAP $f$ on RB $r$ can be expressed as follows:
			\small
			\normalsize
			\begin{align}\label{eq:interference}
			I_{fru}\!=w_{fru}a_{fr}p_{fr}h_{fru}+s_{fru}\zeta(1-a_{fr})p_{fr}h_{fru}+ \sum_{\substack{f^{\prime} \in \mathcal{F}\\
				f^{\prime}\neq f}}p_{f^{\prime}r}h_{f^{\prime}ru}.
			\end{align}
		That is, on one hand, if user $u$ is served by FAP $f$ as strong user on RB $r$ ( i.e., $s_{fru}=1$, then $w_{fru}=0$), then user $u$, first, applies successive interference cancellation (SIC) to remove the corresponding weak user's signal before decoding its own signal. In other words, the strong user, first, decodes the signal for the weak user, removes it from the received signal, then decodes its own signal. $\zeta$ is a factor that represents the imperfection of the SIC. Its value ranges from $0$ to $1$, $\zeta=0$ implies perfect SIC.
		 The interference power is, therefore, the summation of the signal power received from all other FAPs  on RB $r$ (third term of eq.~\eqref{eq:interference}) plus the remaining interference from the weak user's signal due to imperfect SIC (second term of eq.~\eqref{eq:interference}). On the other hand, if user $u$ is served by FAP $f$ as weak user on $r$, i.e., $w_{fru}=1$, then it directly decodes its signal, by treating the signal for the strong user as interference.
Consequently, user $u$ is subjected to the interference from the strong user signal (first term of eq.~\eqref{eq:interference}) and to the interference from all other FAPs.

			The resulting DL rate at user $u$ served by FAP $f$ on RB $r$ is hence:		
			\vspace{-0.2cm}
			\begin{equation}\label{eq:rate}
			C_{fru}(\boldsymbol{p}_{r}, a_{fr},s_{fru},w_{fru})=\beta \log\!\left(1+\frac{\chi_{fru}}{I_{fru}+\beta N_{o}}\right),
			\end{equation}
			\normalsize			
			\noindent where vector $\boldsymbol{p}_{r}=[p_{1r},  \ldots,p_{Fr}]^{\top}\in \mathbb{R}_+^{F \times 1}$
denotes the transmit power vector by all FAPs on RB $r$.  $N_{o}$ is the noise spectral density at the users' receivers, and $\beta$ is the transmission bandwidth for every RB.
The overall DL rate at the user $u$ is
			\begin{equation}\label{eq:rate}
			C_{u}(\boldsymbol{P}, \boldsymbol{A},\boldsymbol{S},\boldsymbol{W})=\sum_{f\in\mathcal{F}}\sum_{r\in\mathcal{R}} C_{fru}(\boldsymbol{p}_{r}, a_{fr},s_{fru},w_{fru}),
			\end{equation}
			\normalsize
			where $\boldsymbol{P}$ and $\boldsymbol{A}$ are matrices of dimension $F\times R$, their $(f,r)$-th elements being, respectively, $p_{fr}$ and $a_{fr} $. $ \boldsymbol{S} $ and $ \boldsymbol{W} $ are matrices of dimension $F\times R \times U$, their $(f,r,u)$-th elements being, respectively, $s_{fru}$ and $w_{fru}$.
			
Let $\psi_{f}$ be the summation of the DL rate of all users served by FAP $f$, $\psi_{f}$ can be expressed as
\begin{equation}\label{eq:DL_fate_FAP_f}
\psi_{f}\!\left(\boldsymbol{P}, \boldsymbol{a}_{f},\boldsymbol{S},\boldsymbol{W}\right)=\sum _{u\in\mathcal{U}}C_{fru}(p_{fr},a_{fr}, s_{fru},w_{fru}).
\end{equation}
$\forall f\in\mathcal{F}$, the vector $\boldsymbol{a}_{f}=[a_{f1},  \ldots,a_{fR}]^{\top}\in \mathbb{R}_+^{R \times 1}$ groups  the power ratios allocated to the strong users of FAP $f$.

  In the sequel, we assume the practical case where there are more than $2R$ users in the network, i.e., $U\geq 2R$.

		\section{Overall Optimization Problem Formulation}\label{sec:OptimizationProblem}
			\subsection{Problem Formulation}\label{subsec:ProblemFormulation}
			The objective of this work is to maximize the weighted sum-rate of the FRAN by optimizing the user assignment, transmit power allocation and NOMA power split ratios, over all FAPs and RBs. Let  $\Theta(\boldsymbol{P}, \boldsymbol{A},\boldsymbol{S},\boldsymbol{W})=\sum_{u\in \mathcal{U}} \alpha_{u}C_{u}(\boldsymbol{P}, \boldsymbol{A},\boldsymbol{S},\boldsymbol{W})$ be the weighted sum-rate function, where the weight $\alpha_{u}$ associated with user $u$ is typically set to maximize the proportional fairness metric (cf. Sec.~\ref{subsec:simulation_setup}).
The optimization problem of interest can then be formulated as follows: 		

				\vspace{-0.4cm}
			\small
			\begin{subequations}\label{eq:Optimization}
				\begin{align}
				\underset{\substack{\boldsymbol{P}, \boldsymbol{A},\boldsymbol{S},\boldsymbol{W}} }{\tr{max}}&\ \Theta(\boldsymbol{P}, \boldsymbol{A},\boldsymbol{S},\boldsymbol{W})\label{eq:Objective_sumlog_1RB}\\
				\tr{s.t.\ \ } 			& 0 \leq a_{fr}\leq 1,\ \forall (f,r)\in \left(\mathcal{F}\times\mathcal{R}\right), \label{eq:powerFraction_1RB}	\\			
				& \sum_{r \in \mathcal{R}}p_{fr}\leq \bar{P}_f,\ \forall f\in \mathcal{F}, \label{eq:powefrudget}	\\								
				& s_{fru},  w_{fru}\in \{0,1\},\ \forall (f,r,u) \in \left(\mathcal{F}\times\mathcal{R}\times\mathcal{U}\right),\label{eq:binary_constraint_1RB}	\\
				& \sum_{u \in \mathcal{U}} 	s_{fru}=1,			\quad \forall (f,r)\in \left(\mathcal{F}\times\mathcal{R}\right),\label{eq:OneStrongUser_1RB}\\
				& \sum_{u\in \mathcal{U}} w_{fru}=	1,	\quad \forall (f,r)\in \left(\mathcal{F}\times\mathcal{R}\right),\label{eq:lOneorLessWeakUser_1RB}\\
				& \sum_{f\in \mathcal{F}}s_{fru}+ w_{fru} \leq	1,	\quad \forall (u,r)\in \left(\mathcal{U}\times\mathcal{R}\right),\label{eq:OneUserperFAP}\\
				&\psi_{f}\!\left(\boldsymbol{P}, \boldsymbol{a}_{f},\boldsymbol{S},\boldsymbol{W}\right)\leq \bar{C_{f}},\  \forall f\in \mathcal{F},	\label{eq:FronthaulConstraint_1}				
			\end{align}
			\end{subequations}
			\normalsize

\noindent where the optimization is over the assignment matrices $\boldsymbol{S}$ and $\boldsymbol{W}$, the transmit power allocation  $\boldsymbol{P}$ over all RBs and FAPs, and the power split ratio $\boldsymbol{A}$ between strong and weak users for all RBs and FAPs.
The first constraint~\eqref{eq:powerFraction_1RB} gives the domain of definition of the NOMA power split ratios $a_{fr}$.
The second constraint ~\eqref{eq:powefrudget} expresses the maximum power budget $\bar{P}_f $ of each FAP $f$, that may be used to transmit over all RBs. Next is the binary constraint~\eqref{eq:binary_constraint_1RB} of assignment variables $s_{fru}$ and $w_{fru}$. Equations \eqref{eq:OneStrongUser_1RB} and \eqref{eq:lOneorLessWeakUser_1RB} constrain every FAP to have, respectively, one strong user and one weak user on each one of its RBs.
The constraint \eqref{eq:OneUserperFAP} is the FRAN-specific constraint, which ensures that every user is served by at most one FAP on a given RB. A user, however, can be served by many FAPs on different RBs by  leveraging the centralized cloud architecture as well.
Lastly,  the summation of the DL rate of all users served by FAP $f$, $\psi_{f}$~\eqref{eq:DL_fate_FAP_f}, is limited {via~\eqref{eq:FronthaulConstraint_1}} to the capacity of the fronthaul link $\bar{C}_{f}$  as in \cite{Dai2015}.

In the expression of the objective function~\eqref{eq:Objective_sumlog_1RB}, each term in the summation is a function of both binary and continuous variables, which complicates the optimization problem. To simplify the presentation, we define $C_{fru}^{(s)}$ ($C_{fru}^{(w)}$) as the rate of user $u$ when it is served by FAP $f$ on RB $r$ as the strong user (the weak user). $C_{fru}^{(s)}$ and $C_{fru}^{(w)}$ can be expressed as follows:
\small
\begin{equation}\label{eq:rateStrong}
C_{fru}^{(s)}(\boldsymbol{p}_{r},a_{fr})\!=\!\beta \log\!\!\left(\!\!1+\frac{ a_{fr}p_{fr}h_{fru}}{\displaystyle\sum_{\substack{f^{\prime}\in \mathcal{F}\\ f^{\prime} \neq f}} \!\!p_{f^{\prime}r} h_{f^{\prime}ru}+\beta N_{o}}\!\!\right),
\end{equation}
and
\small
\begin{align}\label{eq:rateWeak}
&C_{fru}^{(w)}(\boldsymbol{p}_{r},a_{fr})=\beta \log\!\Bigg( 1\ +\!\frac{ (1-a_{fr})p_{fr}h_{fru}}{\!\!\displaystyle\sum_{\substack{f^{\prime}\in \mathcal{F}\\ f^{\prime} \neq f}} \! p_{f^{\prime}r} h_{f^{\prime}ru}+a_{fr}p_{fr}h_{fru}\!+\!\beta N_{o}}\!\Bigg).
\end{align}
\normalsize
Using these two expressions, the objective function~\eqref{eq:Objective_sumlog_1RB} can be re-written as follows:		
			\begin{align}\label{eq:objectiveReformulation}
			\!\!\!\!\Theta(\boldsymbol{P}, \boldsymbol{A},\boldsymbol{S},\boldsymbol{W})&\!=\sum_{f\in \mathcal{F}}\sum_{r\in \mathcal{R}}\sum_{u\in \mathcal{U}}\!\!\left(s_{fru}\alpha_{u}C_{fru}^{(s)}(\boldsymbol{p}_{r},a_{fr})+w_{fru}\alpha_{u}C_{fru}^{(w)}(\boldsymbol{p}_{r},a_{fr})\right).
			\end{align}
			\normalsize
The fronthaul link capacity constraint~\eqref{eq:FronthaulConstraint_1} can be also re-written as follows:
\begin{equation}
\psi_{f}\!\left(\boldsymbol{P}, \boldsymbol{a}_{f},\boldsymbol{S},\boldsymbol{W}\right)=\sum_{r\in \mathcal{B}}\sum_{u\in \mathcal{U}}\!\!\left(s_{fru}C_{fru}^{(s)}(\boldsymbol{p}_{r},a_{fr})+w_{fru}C_{fru}^{(w)}(\boldsymbol{p}_{r},a_{fr})\right).
\end{equation}
In the above expressions of the objective function and the fronthaul link capacity constraints, one can observe that the rate terms~\eqref{eq:rateStrong} and~\eqref{eq:rateWeak} are no longer functions of the binary variables. Such observation justifies the rationale behind iteratively solving the binary part and the continuous part of problem~\eqref{eq:Objective_sumlog_1RB} in a separate fashion, as adopted in the remaining parts of the paper.
		
\subsection{Problem Analysis and Overview of the Proposed Solution}

			The optimization problem defined in \eqref{eq:Objective_sumlog_1RB} is a mixed-integer optimization problem, i.e., some variables are binary $(\boldsymbol{S},\boldsymbol{W})$ and some are continuous $(\boldsymbol{P}, \boldsymbol{A})$. Such problems are generally hard to solve, particularly problem~\eqref{eq:Optimization}, which cannot be globally solved in polynomial time. Such difficulty is due to {the mutual interference between the FAPs, which interweaves the assignment, transmit power allocation and NOMA power splitting problems. On one hand, the optimal user assignment depends} on the transmit power allocation and NOMA power split ratios within each FAP $(\boldsymbol{P}, \boldsymbol{A})$ which on the other hand depend on the user assignment ($\boldsymbol{S}$ and $\boldsymbol{W}$). In addition, the continuous part of the optimization problem defined in~\eqref{eq:Optimization} is not convex, due to the non-convexity of the objective function and the fronthaul capacity constraint~\eqref{eq:FronthaulConstraint_1}.

			 Based on the structure of the problem and the network architecture, this paper proposes an iterative algorithm that efficiently solves problem~\eqref{eq:Optimization}. A brief description of the overall algorithm is given in Algorithm~\ref{alg:overallAlgorithm}, where $x^{(i)}$ denotes the value of $x$ at the $i$-th iteration. The algorithm stops when the increase in objective function value is less than $\epsilon$. The algorithm starts by fixing the starting points $\boldsymbol{P}^{(o)}$ and $\boldsymbol{A}^{(o)}$, and the value of $\epsilon$ which determines the exit condition.
			 Each iteration of the algorithm consists of three steps: first, solving the optimization problem for the binary variables $\boldsymbol{S}$ and $\boldsymbol{W}$ given fixed $\boldsymbol{P}$ and $\boldsymbol{A}$ (solved in details in Sec..~\ref{sec:AssignmentAlgorithm}). At the end of this step, each RB at every FAP is assigned to a pair of strong and weak users. The second and third steps are solving the problem for its continuous variables $\boldsymbol{P}$ and $\boldsymbol{A}$ given the assignment found in the first step. The second step determines how the  transmit power budget $\bar{P}_{f}$ available at every FAP is allocated to its RBs, i.e., find $\boldsymbol{P}$ (solved in details in Sec.~\ref{sec:PowerAllocation} ). The final step determines how, given $\boldsymbol{P}$,  the transmit power allocated on each RB is split between its strong and weak users, i.e., find $\boldsymbol{A}$ (solved in details in Sec.~\ref{sec:PowerSplitOptimization} ). This paper shows that for this last step, the NOMA power split optimization problem is separable in $f$'s (cf. Sec.~\ref{sec:PowerSplitOptimization}). Therefore, it can be separately solved at every FAP instead of requiring a centralized processing at the BBU pool. This is well suitable to the FRAN principle where more intelligence and control can be pushed towards the network edge, thereby alleviating the fronthaul traffic burden and meeting the requirements of delay-stringent real-time applications.
		
Each step of the algorithm is carefully detailed in the following sections.

			\begin{algorithm}
				\small
				\begin{algorithmic}[1]
				\State{Set $\epsilon$, $\boldsymbol{P}^{(o)}$ and $\boldsymbol{A}^{(o)}$}
				\State{$i=1$.}
				\Do{ }
					\State{Assignment: find $\boldsymbol{S}^{(i)}$ and $\boldsymbol{W}^{(i)}$, for fixed $\boldsymbol{P}^{(i-1)}$ and $\boldsymbol{A}^{(i-1)}$ (Sec.~\ref{sec:AssignmentAlgorithm}).}
					\State{Transmit power allocation: find $\boldsymbol{P}^{(i)}$, for fixed $\boldsymbol{S}^{(i)}$, $\boldsymbol{W}^{(i)}$ and $\boldsymbol{A}^{(i-1)}$  (Sec.~\ref{sec:PowerAllocation}).}
					\State{NOMA Power split optimization: find $\boldsymbol{A}^{(i)}$, for fixed $\boldsymbol{S}^{(i)}$, $\boldsymbol{W}^{(i)}$ and $			\boldsymbol{P}^{(i)}$  (Sec.~\ref{sec:PowerSplitOptimization}).}
					\State{$i=i+1.$}
					\doWhile{$ f\!\left(\boldsymbol{P}^{(i)}, \boldsymbol{A}^{(i)},\boldsymbol{S}^{(i)},\boldsymbol{W}^{(i)}\right)- f\!\left(\boldsymbol{P}^{(i-1)}, \boldsymbol{A}^{(i-1)},\boldsymbol{S}^{(i-1)},\boldsymbol{W}^{(i-1)}\right)\geq \epsilon$.}		
				\end{algorithmic}
				\caption{Overall algorithm: general description}\label{alg:overallAlgorithm}
			\end{algorithm}
			\normalsize

\section{ Proposed Assignment Methods}\label{sec:AssignmentAlgorithm}		
			This section describes the first step of Algorithm~\ref{alg:overallAlgorithm} where one weak and one strong user are optimally assigned to each RB of every FAP by finding the optimal $\boldsymbol{S}$ and $\boldsymbol{W}$ given fixed power allocation $\boldsymbol{P}$ and fixed NOMA power split $\boldsymbol{A}$.  For fixed $\boldsymbol{P}$ and $\boldsymbol{A}$, problem~\eqref{eq:Optimization} boils down to a binary optimization problem, formulated as follows:
			\vspace{-0.3cm}
				\begin{subequations}\label{eq:binaryOptimization}
				\begin{align}
				\underset{\substack{\boldsymbol{S},\boldsymbol{W}} }{\tr{max}}&\ \sum_{\substack{f\in \mathcal{F}\\ r\in \mathcal{R}\\ u\in \mathcal{U}}} s_{fru}\alpha_{u}C_{fru}^{(s)}(\boldsymbol{p}_{r},a_{fr})+w_{fru}\alpha_{u}C_{fru}^{(w)}(\boldsymbol{p}_{r},a_{fr})\\
\tr{s.t.\ \ } 			& \eqref{eq:binary_constraint_1RB},\eqref{eq:OneStrongUser_1RB},\eqref{eq:lOneorLessWeakUser_1RB},\eqref{eq:OneUserperFAP}	\  \text{and}\ \eqref{eq:FronthaulConstraint_1}.
				\end{align}
				\end{subequations}
				
To solve this problem, we propose two assignment algorithms: the first one is based on the Hungarian algorithm  (detailed in Sec.~\ref{Subsec:HungarianAssignment}).
The second one is based on the Multiple Choice Knapsack Problem (MCKP) algorithm (detailed in Sec.~\ref{subsec:Knapsack}).	

\subsection{Hungarian Based Assignment Algorithm}\label{Subsec:HungarianAssignment}
As previously mentioned, the goal of this step is to find the optimal user assignment $\boldsymbol{S}$ and $\boldsymbol{W}$ given fixed $\boldsymbol{P}$ and $\boldsymbol{A}$. In that case, it can be observed from~\eqref{eq:Optimization} that, the user assignment is tangled between the RBs, only because of the fronthaul capacity constraint~\eqref{eq:FronthaulConstraint_1}. In other words, without the fronthaul capacity constraint~\eqref{eq:FronthaulConstraint_1}, the overall utility function is the summation of the utilities from each RB and the problem can be optimally solved by separately solving for each RB:
\vspace{-0.7cm}
\begin{subequations}\label{eq:hungarianAssignment}
		\begin{align}
				\underset{\substack{\boldsymbol{S}_{r},\boldsymbol{W}_{r}} }{\tr{max}}&\ \sum_{\substack{f\in \mathcal{F}\\ u\in \mathcal{U}}} s_{fru}\alpha_{u}C_{fru}^{(s)}+w_{fru}\alpha_{u}C_{fru}^{(w)}\\
				\tr{s.t.\ \ }								& s_{fru},  w_{fru}\in \{0,1\},\ \forall (f,u) \in \left(\mathcal{F}\times\mathcal{U}\right),\\
				& \sum_{u \in \mathcal{U}} 	s_{fru}=1,			\quad \forall f\in \mathcal{F},\\
				& \sum_{u\in \mathcal{U}} w_{fru}=	1,	\quad \forall f\in \mathcal{F},\\
				& \sum_{f\in \mathcal{F}}s_{fru}+ w_{fru} \leq	1,	\quad \forall u\in \mathcal{U}.
		\end{align}
\end{subequations}
The optimization variables are $\boldsymbol{S}_{r}$ and $\boldsymbol{W}_{r}$, two binary matrices of dimension $F\times U$ for all assignment variables corresponding to RB $r$. As shown in the problem formulation above~\eqref{eq:hungarianAssignment}, the fronthaul constraint is rather discarded so as to enable utilizing the classical Hungarian method. The fronthaul link capacity constraint~\eqref{eq:FronthaulConstraint_1} is, however, reinforced while adjusting the transmission power in the subsequent step \footnote{The MCKP-based user assignment algorithm proposed in Sec.~\ref{subsec:Knapsack} satisfies the fronthaul constraints.}.

	For the optimization problem defined in~\eqref{eq:hungarianAssignment}, the contribution of every FAP to the utility function is the summation of the weak user rate and the strong user rate, which are known for fixed $\boldsymbol{P}$ and $\boldsymbol{A}$. It can be thus assumed that, for a given RB, each FAP is equivalent to two separate virtual FAPs, each serving a unique user. Moreover, a user can only be served by one virtual FAP for a given RB.  The problem described in~\eqref{eq:hungarianAssignment} is, therefore, transformed to a one user-to-one virtual FAP assignment problem. That is, for a given RB, a regular one-to-one optimal assignment algorithm can be applied to solve~\eqref{eq:hungarianAssignment}, while satisfying all the constraints. There are several well-known combinatorial optimization algorithms that solve such one-to-one assignment problems in the literature. In this paper, we choose the Hungarian algorithm for its relative implementation simplicity and polynomial execution time \cite{Kuhn1955}.
			Let $K_{r}$ be the $U \times 2R$ cost matrix associated with the considered assignment problem. The $(u,r)$-th and $(u,R+r)$-th elements of $K_{r}$ are the additive inverse of the utility that {results} from assigning user $u$ to be the strong  and weak user of the  $f$-th FAP's $r$-th RB, respectively, i.e., the $(u,r)$-th and $(u,R+r)$-th elements of $K_{r}$ are $-C_{fru}^{(s)}$ and $-C_{fru}^{(w)}$, respectively (i.e., turn the utility terms into cost terms, and minimize the resulting objective). A detailed description of the proposed assignment is given in Algorithm~\ref{alg:assignmentAlgorithm}.
				\begin{algorithm}
				\small
				\begin{algorithmic}[1]
				\State{Input: $\boldsymbol{P}^{(i-1)} $, $\boldsymbol{A}^{(i-1)}$}
					\For {$r=1\rightarrow R$}
					\State{Construct $K_{r}$}
					\State{Find $\boldsymbol{S}_{r}^{(i)}$, $\boldsymbol{W}_{r}^{(i)}$}\Comment {Apply Hungarian algorithm to $K_{r}$.}
					\EndFor	
					\State{Output: $\boldsymbol{S}^{(i)}, \ \boldsymbol{W}^{(i)} $.}
				\end{algorithmic}
				\caption{Hungarian-based Assignment Algorithm}\label{alg:assignmentAlgorithm}
			\end{algorithm}
			\normalsize
			This algorithm has to be performed at the cloud BBUs as it is not separable per FAP.
		 For weighted sum-rate maximization, the result does not depend on the order of RBs, i.e., the assignment can be performed in parallel.
		 			
			\subsection{Auction-MCKP-Based Assignment Algorithm}\label{subsec:Knapsack}
			
			
As previously mentioned, one issue of the Hungarian-based assignment algorithm is that it does not account for the fronthaul link capacity constraint~\eqref{eq:FronthaulConstraint_1}. Hence, the applicability of the resulting user-to-RB assignment solution relies on the subsequent power allocation step, which adjusts the transmission power on each RB such that the fronthaul link capacity constraint is satisfied. Instead, we now develop an alternative assignment method that directly accounts for the fronthaul capacity, and hence, can be applied independently from the power allocation step.

Considering one FAP, the assignment problem can be formulated as a Multiple-Choice Knapsack Problem (MCKP). MCKP consists of packing a number of items in a limited capacity knapsack, where the items to be packed belong to disjoint classes of items, and each item has a certain profit and weight \cite{Kellerer2004}. The goal is then to choose exactly one item from each class such that the profit/utility sum is maximized without exceeding the knapsack's capacity. In our case, we consider that each FAP $f$ is a knapsack with capacity $\bar{C}_f$ that has to be filled with $R$ pair of users (items), i.e., one pair from every RB (class).

The complication here is that our target problem cannot be solved independently for every FAP, due to constraint~\eqref{eq:OneUserperFAP}. More precisely, a user cannot be served by two different FAPs on the same RB, i.e., a FAP cannot choose a pair of users that includes users chosen by other FAPs for the same class (RB $ r$). This paper, therefore, proposes solving this conflict by an auction approach where every FAP bids by proposing a price for every user chosen by the initial MCKP-greedy algorithm.
At the end of every bidding iteration, the price of each user for a given RB is set to the maximum of all received bids. This bidding process is repeated until the prices of all users do not change.

This subsection is organized as follows. First, the notations used in this substep of the algorithm are presented in Sec.~\ref{subsec:Notation}. Then, we detail how MCKP-greedy algorithm is applied to allocate users to every FAP in Sec.~\ref{subsec:greedy}. Lastly, Sec.~\ref{subsec:auction} presents the overall assignment algorithm where auction algorithm is used to solve the conflicts among FAPs.

%
\subsubsection{Notations}\label{subsec:Notation}
Considering the $f$th FAP, let $\mathcal{O}_{fr}$ be the set of all possible combinations of users on RB $ r $.
The $i$-th item in $\mathcal{O}_{fr}$, called herein $\mathcal{O}_{fr}(i)$, denotes the user pair $(u_{s}(i),u_{w}(i))$, where $u_{s}(i)$ is the strong user and $u_{w}(i)$ is the weak user. The utility corresponding to the $i$-th item in $\mathcal{O}_{fr}$  is:
\begin{equation}\label{eq:utilityForMCKP}
\upsilon_{fr}(i)=\alpha_{u_{s}(i)}C_{fru_{s}(i)}^{(s)}+\alpha_{u_{w}(i)}C_{fru_{w}(i)}^{(w)}-\rho_{r}^{(t)}(u_{s}(i))-\rho_{r}^{(t)}(u_{w}(i)),
\end{equation}
where the arguments of the functions $ C_{fru_{s}(i)}^{(s)} $and $ C_{fru_{w}(i)}^{(w)}$ (originally defined in~\eqref{eq:rateStrong} and~\eqref{eq:rateWeak}) are omitted in~\eqref{eq:utilityForMCKP} to simplify the presentation, and where $\rho_{r}^{(t)}(u_{s}(i))$ and $\rho_{r}^{(t)}(u_{w}(i))$ are the prices of the users $u_{s}(i) $ and $u_{w}(i)$ at the $t$-th iteration. Note that the prices of a user are different  across different RBs.
The weight (cost) corresponding to the $i$-th item in $\mathcal{O}_{fr}$ is:
\begin{equation}
c_{fr}(i)=C_{fru_{s}(i)}^{(s)}+C_{fru_{w}(i)}^{(w)}.
\end{equation}
Let $\rho_{fr}(u)$ be the price that FAP $f$ bids for user $u$ on RB $r$.

For two items $i$ and $j$ in $\mathcal{O}_{fr}$, item $j$ is \textbf{dominated} by item $i$ if
\begin{equation}
c_{fr}(i)\leq c_{fr}(j) \text{ and } \upsilon_{fr}(i)\geq\upsilon_{fr}(j).
\end{equation}
In other words, an element is \textbf{dominated} if there exists another element with higher or equal profit, but less weight. Therefore, a dominated element should not be in the set of items to choose from.

The incremental profit  between two elements $i$ and $j$ is a measure of how much utility a FAP gains if it chooses item $i$ instead of item $j$ from $\mathcal{O}_{fr}$.
The incremental weight has a similar interpretation (how much more weight is added to a FAP if it chooses item $i$ instead of item $j$ from $\mathcal{O}_{fr}$). We define the incremental efficiency between two elements $i$ and $j$ as the ratio  between their incremental profit and incremental weight, i.e.,
\begin{equation}
\tilde{e}_{fr}(j,i)=\frac{\upsilon_{fr}(j)-\upsilon_{fr}(i)}{c_{fr}(j)-c_{fr}(i)}.
\end{equation}

For three items $i$, $j$ and $k$ in $\mathcal{O}_{fr}$, with $c_{fr}(i)<c_{fr}(j)<c_{fr}(k) $ and  $\upsilon_{fr}(i)<\upsilon_{fr}(j)<\upsilon_{fr}(k) $,  we say that $j$ is \textbf{LP-dominated} (linear programming dominated) by $i$ and $k$ \cite{Kellerer2004}  if
\begin{equation}\label{eq:incrementalEfficiency}
\tilde{e}_{fr}(k,j)\geq\tilde{e}_{fr}(j,i)\text {, i.e., }\frac{\upsilon_{fr}(k) -\upsilon_{fr}(j)}{c_{fr}(k) -c_{fr}(j)}\geq \frac{\upsilon_{fr}(j) -\upsilon_{fr}(i)}{c_{fr}(j) -c_{fr}(i)}.
\end{equation}
In other words, item $j$ is \textbf{LP-dominated} by $i$ and $k$ if the incremental efficiency between $k$ and $j$ is larger than the one between $j$ and $i$ \cite{Kellerer2004}.

Let  $\mathcal{E}_{fr}$ be the set of \textbf{LP-extreme} items corresponding to $\mathcal{O}_{fr}$. This set is obtained by eliminating all dominated and LP-dominated elements in $\mathcal{O}_{fr}$.		

\subsubsection{MCKP-Greedy Algorithm}\label{subsec:greedy}
Based on the above notations, the paper now details how $R$ pairs of users are assigned to each FAP $f$. As previously mentioned, the goal for FAP $f$ is to choose one pair of users (item) such that every RB (class) maximizes its utility function without exceeding the fronthaul capacity. To this end, FAP $f$ first constructs $R$ classes of items ($\mathcal{O}_{fr} \text{ for } r=1,\ldots,R$).
Each class contains all possible combinations of two users (a pair of strong $(u_{s})$ and weak $(u_{w})$ users). For each pair of users (each item), the user with better channel is the strong user, while the one with worse channel is the weak user. Therefore, there are ${U\choose 2}$ items in every class. Each item is characterized with a profit (the utility value added to the overall utility function if the corresponding user pair is chosen) and a weight (the capacity that the corresponding user pair needs in the fronthaul link).
The goal is then to choose the best pair from each set $\mathcal{O}_{fr}$  ($r=1,\ldots,R$) such that the utility is maximized and the fronthaul capacity is not exceeded. Let $x_{fr}(i)$ be a binary variable that is equal to 1 if the $i$-th item in $\mathcal{O}_{fr}$ is chosen. In this part, constraint~\eqref{eq:OneUserperFAP} is discarded, as it is the object of the auction algorithm detailed in the next subsection. Using the notations introduced in Sec.~\ref{subsec:Notation}, the binary optimization for FAP $f$ can be reformulated as follows:
\vspace{-0.3cm}
\begin{subequations}\label{eq:binaryOptimizationKnapsack}
				\begin{align}
				\underset{x_{fr}, r \in \mathcal{R} }{\!\!\!\!\!\!\!\!\tr{max}}&\ \sum_{\substack{r\in \mathcal{R}}}\sum_{i=1}^{\vert \mathcal{O}_{fr} \vert } x_{fr}(i)\upsilon_{fr}(i)\label{eq:knapsackUtility}\\				
				\tr{s.t.\ \ }								& \sum_{\substack{r\in \mathcal{R}}}\sum_{i=1}^{\vert \mathcal{O}_{fr} \vert } x_{fr}(i)c_{fr}(i)\leq \bar{C}_{f}\label{eq:fronthaul}\\
				& \sum_{i=1}^{\vert \mathcal{O}_{fr} \vert } x_{fr}(i)=1,\ \forall r\in\mathcal{R},\label{eq:one_item_per_class}\\
				& x_{fr}(i)\in \{0,1\},\ \forall r\in\mathcal{R},\ \forall i\in\{1,\ldots,\vert\mathcal{O}_{fr}\vert \}\label{eq:integrality_constraint},
				\end{align}
				\end{subequations}
The objective function~\eqref{eq:knapsackUtility} is the total utility at FAP $f$, which is the summation of the utilities of chosen pairs~\eqref{eq:knapsackUtility}. At the beginning of the algorithm, the utility associated to every user pair is equal to the summation of the utilities of the strong and weak users forming the user pair. As the algorithm progresses, the user pair's utility includes the price of the user pair from the auction's previous iteration as defined in~\eqref{eq:utilityForMCKP}. The fronthaul capacity constraint for FAP $ r $, equivalent to~\eqref{eq:FronthaulConstraint_1}, is expressed in~\eqref{eq:fronthaul}. The constraint~\eqref{eq:one_item_per_class} forces FAP $f$ to choose one pair from each class (RB $r$). Therefore, constraint~\eqref{eq:one_item_per_class} is equivalent to \eqref{eq:OneStrongUser_1RB} and  \eqref{eq:lOneorLessWeakUser_1RB} in the original binary optimization problem. The integrality constraint on the binary variables $x_{fr}(i)$ is expressed in~\eqref{eq:integrality_constraint}.
Note that, for the MCKP to be feasible, we assume that, for every FAP $f$:
\begin{equation}\label{eq:MCPKPcondition1}
\sum_{r=1}^{R} \underset{\substack{i \in \mathcal{O}_{fr}} }{\tr{min}} c_{fr}(i) \leq \bar{C}_{f},
\end{equation}
and for every item (user pair) $i \in \mathcal{O}_{fr}$,
\begin{equation}\label{eq:MCPKPcondition2}
c_{fr}(i) +\sum_{\substack{r^{\prime}=1\\r^{\prime}\neq r}}^{R} \underset{\substack{i \in \mathcal{O}_{fr^{\prime}}} }{\tr{min}} c_{fr^{\prime}}(i)\leq \bar{C}_{f}.
\end{equation}

By relaxing the integrality constraint~\eqref{eq:integrality_constraint}, the MCKP in~\eqref{eq:binaryOptimizationKnapsack} becomes a linear MCKP which can be solved optimally using the MCKP-greedy algorithm \cite{Kellerer2004}.
The MCKP-greedy algorithm is based on the transformation of the MCKP to a similar problem corresponding to one instance of Knapsack Problem (KP). Each instance is then solved optimally using the KP-greedy algorithm \cite{Kellerer2004} (i.e., by adding an element to the sack in decreasing order of efficiency). The solution of the original problem is then constructed from the solution of the linear relaxation.
The details of the MCKP-greedy algorithm that solves~\eqref{eq:binaryOptimizationKnapsack} are given in Alg.~\ref{alg:KnapsackAssignment}, and explained as follows.

First, for every RB $r$, the FAP constructs a class (set $\mathcal{O}_{fr}$) containing all the possible combinations of two users, and their corresponding utilities and weights (line 2). The elements in each class are sorted according to increasing weight (line 3). Then, for every class, the LP-extreme set $\mathcal{E}_{fr}$  is constructed by eliminating the dominated and LP-dominated elements from the ordered set (line 4). Next, the incremental efficiency between every pair of successive elements in $\mathcal{E}_{fr}$ is determined as follows:
\begin{equation}
\tilde{e}_{fr}(i,i-1)=\frac{\upsilon_{fr}(i)-\upsilon_{fr}(i-1)}{c_{fr}(i)-c_{fr}(i-1)}.
\end{equation}



The items from all $\mathcal{E}_{fr}$ sets ($r=1,\ldots,R$) are placed in a new set $\boldsymbol{\mathcal{E}}_{f}$ in decreasing order of incremental efficiency (line 12). The algorithm, then, adds items to the sack in that order, until the capacity of the sack is exceeded.
The output of the MCKP-greedy algorithm is $\boldsymbol{i}^{*}_{f}=[i^{*}_{f1},\ldots,i^{*}_{fR}]$, i.e., the index of every chosen user pair from each class $r$ in the set $\mathcal{E}_{fr}$ for FAP $f$. Conditions~\eqref{eq:MCPKPcondition1} and~\eqref{eq:MCPKPcondition2} ensure that there is a chosen pair from every class.

			\begin{algorithm}
				\small
				\begin{algorithmic}[1]
				\For {every class (RB) $r$}
							\State{Construct the set $\mathcal{O}_{fr}$ and the utility and weight vectors $\upsilon_{fr}$ and  $c_{fr}$}
							\State{Sort elements in $\mathcal{O}_{fr}$ according to increasing weight $c_{fr}$}
							\State{Construct the set of LP-extreme items $\mathcal{E}_{fr}$ from $\mathcal{O}_{fr}$}	
							\State{Calculate the incremental efficiency between every two successive element in $\mathcal{E}_{fr}$, i.e., calculate $\tilde{e}_{fr}(i,i-1)$ for $i=[2,..,\vert \mathcal{E}_{fr}\vert]$}
					
					\EndFor	
				\State{$\boldsymbol{\mathcal{E}}_{f}$= items from all $\mathcal{E}_{fr}$ sorted  according to decreasing $\tilde{e}_{fr}$}		
					\State{Initialize $\tilde{C}_{f}=C_{f}-\sum_{r=1}^{R} c_{fr}(1)$}
					\Do{ }
					\State{Add next user pair from $\boldsymbol{\mathcal{E}}_{f}$}
					\State{\textit{(Assuming the item is from $\mathcal{E}_{fr}$ and its index is $i$)}}
					\State{$\tilde{C}_{f}=\tilde{C}_{f}-\tilde{c}_{fr}(i)$}
					\If {$\tilde{C}_{f}\geq 0$}
							\State	{$i^{*}_{fr}=i$}
   				 	\EndIf
					\doWhile{$ \tilde{C}_{f}>0$.}		
		\State{Output: $\boldsymbol{i}^{*}_{f}=[i^{*}_{f1},\ldots,i^{*}_{fR}]$.}
				\end{algorithmic}
				\caption{MCKP algorithm, $f$-th FAP}\label{alg:KnapsackAssignment}
			\end{algorithm}
			\normalsize


\subsubsection{Overall MCKP-Auction Based Assignment Algorithm}		\label{subsec:auction}
By  independently applying~\ref{subsec:greedy}, every FAP has $2R$ tentatively assigned users.
The overall assignment algorithm, however, must satisfy  constraint~\eqref{eq:OneUserperFAP}. In other words, no user can be served by two different FAPs on the same RB. To overcome this issue, we propose the auction-MCKP-based assignment algorithm detailed in Algorithm~\ref{alg:MCKP_auction_assignment}.
Let $\boldsymbol{\bar{f}}$ be a vector containing the indexes of FAPs that need to bid in the next iteration of the auction. At the initialization, all users' prices are set to zero and all FAPs need to bid, i.e., $\boldsymbol{\bar{f}}=[1;\ldots;F]$ (line 2).

At the beginning of every iteration, every FAP finds the users to bid for each RB $(\boldsymbol{i}^{*}_{f})$ by applying the MCKP-greedy algorithm (line 5). For every chosen user (each user in $i^{*}_{fr}$), the bidding price is set to
\begin{subequations}\label{eq:price_update}
\begin{align}
						\rho_{fr}(u_{s}(i^{*}_{fr}))&=\rho_{fr}(u_{s}((i^{*}_{fr}))+\left(\upsilon_{fr}(i^{*}_{fr})-\bar{\upsilon}_{fr}\right)/2\\
						\rho_{fr}(u_{w}(i^{*}_{fr}))&=\rho_{fr}(u_{w}(i^{*}_{fr}))+\left(\upsilon_{fr}(i^{*}_{fr})-\bar{\upsilon}_{fr}\right)/2,
\end{align}
\end{subequations}
for the strong and weak user, respectively.
That is, the price of a user is set to its old price plus half of the utility that the FAP would loose if it looses the bid (the utility loss for losing a user is half of the utility loss for the user pair). In particular, the utility loss for the best user pair is defined as the difference between its utility ($\upsilon_{fr}(i^{*}_{fr})$) and that of the second best pair from the same class, denoted as $\bar{\upsilon}_{fr}$ (lines 6-8).

After all FAPs in $\boldsymbol{\bar{f}}$ place their bids for every RB $r$, the absolute price of every user $\rho_{r}(u)$ is set to the maximum bid placed for that user on RB $r$ (lines 6-8), i.e.,
\begin{equation}\label{eq:priceAgglomeration}
\rho_{r}^{(t+1)}(u)=\underset{f}{\text{max}}\rho_{fr}(u),\ \forall r \in \mathcal{R}.
\end{equation}
For every RB $r$, the FAP with the highest bid wins the user. Hence, FAP $f$ looses the weak/strong user of RB $r$ if
\begin{align}
					&\rho_{fr}(u_{s}(i^{*}_{fr}))\leq\rho_{r}^{(t+1)}(u_{s}(i^{*}_{fr}))~\text{OR}\  \rho_{fr}(u_{w}(i^{*}_{fr}))\leq\rho_{r}^{(t+1)}(u_{w}(i^{*}_{fr})).
\end{align}
The indexes of FAPs that loose a user on at least one of their RBs are grouped in $\boldsymbol{\bar{f}}$ (the list of bidding FAPs in following iterations). It is worth mentioning that there is no guarantee that the winning FAP at a given iteration would keep that user until the end of the algorithm, as it may loose that user in a subsequent iteration if another FAP bids higher.
 The algorithm continues until all RBs of all FAPs are assigned, i.e., the absolute price of all users on all RBs does not change anymore (line 13). In other words, there are no longer any FAPs placing bids. A proof of convergence for the auction algorithm is provided in \cite{Luo2015}. The last step of the algorithm converts the indexes of the final users for every FAP $\boldsymbol{i}^{*}_{f}$ into the binary variables (lines 14-16):
\begin{equation}
 s_{fru_{s}(i^{*}_{fr})}=1, \ w_{fru_{w}(i^{*}_{fr})}=1 \ \forall (f,r) \in (\mathcal{F}\times\mathcal{R})
 \end{equation}

		\begin{algorithm}
				\small
				\begin{algorithmic}[1]
				\State{Input: $\boldsymbol{P}$, $\boldsymbol{A}$}
				\State{Initialization $t=0$, $ \rho_{fr}(u)=0\ \forall (f,r,u)\in (\mathcal{F}\times \mathcal{R}\times\mathcal{U}))$,  $\rho_{r}^{(0)}(u)=0,\ \forall (u,r)\in (\mathcal{U}\times \mathcal{R}))$, $\boldsymbol{\bar{f}}=[1;\ldots;F]$,  $\boldsymbol{S}=\boldsymbol{0}, \ \boldsymbol{W}=\boldsymbol{0} $.}				
				\Do
					\For {every FAP that needs to bid, i.e., $f \in \boldsymbol{\bar{f}}$}
					\State{Find the list of chosen pairs of users $ \boldsymbol{i}^{*}_{f}=[i^{*}_{f1},\ldots,i^{*}_{fR}]$ by applying alg.~\ref{alg:KnapsackAssignment}}
					
					\For {every RB, $r=1 \rightarrow R$}
						\State{Set the bidding price of the chosen strong and weak users according to eq.~\eqref{eq:price_update} }
					\EndFor	
					\EndFor	
					\State{Price agglomeration, eq.~\eqref{eq:priceAgglomeration}.}
					\State{Construct the list of FAPs that lost at least a user $\boldsymbol{\bar{f}}$}					
					\State Increment the number of iteration, $t=t+1$.				
			 	\doWhile{$\exists (r,u) \text{ such that } \rho_{r}^{(t-1)}(u) \neq \rho_{r}^{(t)}(u) $.}
			 	\For {each RB of every FAP: $f=1 \rightarrow F$, $r=1 \rightarrow R$}					
						\State{Set the assignment variables corresponding to the chosen users to 1.}
					\EndFor					
			 	\State{Output: $\boldsymbol{S}, \ \boldsymbol{W} $.}
				\end{algorithmic}
				\caption{MCKP-Auction Based Assignment}\label{alg:MCKP_auction_assignment}
			\end{algorithm}
			\normalsize

As already pointed out, unlike the Hungarian-based assignment algorithm, the MCKP-based assignment algorithm results satisfies the fronthaul capacity constraint. Therefore, the MCKP-based assignment algorithm can be applied alone without optimized power allocation (cf. Sec.~\ref{subsec:algo_performance}).

\section{Proposed Power Allocation Method}\label{sec:PowerAllocation}
This section details the second step in Algorithm~\ref{alg:overallAlgorithm}, as it focuses on solving the power allocation step. For the assignment solution found in the first step, the available power at every FAP is optimally assigned to its RBs, i.e.,  $\boldsymbol{P}$ is optimized for fixed $\boldsymbol{S}$, $\boldsymbol{W}$ and $\boldsymbol{A}$.
The overall utility function is separable across the RBs, i.e., the overall utility is the summation of the contribution from every RB, cf. eq.~\eqref{eq:objectiveReformulation}. Moreover, the contribution of each RB to the overall utility function depends only on the power allocated to that given RB across all FAPs.
Thus, the power allocation optimization can be formulated as follows:
\small
		\begin{subequations}\label{eq:powerAllocationProblem}
				\begin{align}
				\underset{\substack{\boldsymbol{P}} }{\tr{max}}&\ \Theta(\boldsymbol{P}, \boldsymbol{A},\boldsymbol{S},\boldsymbol{W})=\sum_{\substack{r\in \mathcal{R}}}\left( \sum_{\substack{f\in \mathcal{F}\\ u\in \mathcal{U}}}s_{fru}\alpha_{u}C^{(s)}_{fru}(\boldsymbol{p}_{r},a_{fr})+w_{fru}\alpha_{u}C^{(w)}_{fru}(\boldsymbol{p}_{r},a_{fr})\right)\label{eq:power_objective_separation}\\
				\tr{s.t.\ \ } 	&\sum_{r \in \mathcal{R}}p_{fr}\leq \bar{P}_f,\ \forall f\in \mathcal{F}, \label{eq:powerConstraint1}\\
				&\psi_{f}\!\left(\boldsymbol{P}, \boldsymbol{a}_{f},\boldsymbol{S},\boldsymbol{W}\right)\leq \bar{C_{f}},\  \forall f\in \mathcal{F}.	\label{eq:powerConstraint2}					
				\end{align}
		\end{subequations}
		\normalsize

		The optimization, however, cannot be separately solved for every RB, due to the total power budget and the fronthaul capacity limit for every FAP. 
The separability of the utility function per RB as shown in~\eqref{eq:power_objective_separation}, however, allows us to use ADMM \cite{Wang2017} to solve the power allocation at the BBU pool, as it is well adapted to the structure of our problem as argued in the sequel. Moreover, ADMM provides a relatively simple computational complexity solution as shown in Sec.~\ref{sec:ComplexityAnalysis}. ADMM consists of sequentially solving the optimization for each direction of the optimization variable (in our case, each direction is the power vector on each RB, i.e., $\boldsymbol{p}_r$) while fixing the other variables.
While ADMM is usually applied to linear equality constraints, the constraints in our case are neither linear nor convex. We hence transform the inequality constraints into equality constraints by adding a component-wise maximum function ($g_{1f}$, $g_{2f}$ below) to each inequality constraint~\cite{Giesen2016},  which ensures that an inequality constraint is only considered whenever active.
Note that a constraint is active if its corresponding maximum function ($g_{1f}$ or $g_{2f}$) is positive for the current value of the variables.
The optimization problem~\eqref{eq:powerAllocationProblem} is thus reformulated as follows:

				\begin{subequations}\label{eq:powerAllocationProblem}
				\begin{align}
				\underset{\substack{\boldsymbol{P}} }{\tr{min}}&\ -\Theta(\boldsymbol{P}, \boldsymbol{A},\boldsymbol{S},\boldsymbol{W})\\
				\tr{s.t.\ \ } 	& g_{1f}\!\left(\boldsymbol{p}_{(f)}\right)=0\label{eq:constraint1},\ \forall f\in \mathcal{F},\\
				&g_{2f}\!\left(\boldsymbol{P}\right)=0\label{eq:constraint2},\  \forall f\in \mathcal{F},
				\end{align}
		\end{subequations}
		where the vector $\boldsymbol{p}_{(f)}=[p_{f1},\ldots,p_{fR}]^{\top} \in \mathbb{R}_+^{R \times 1}$ groups the RBs' transmit powers of FAP $f$. We define $g_{1f}$ and $g_{2f}$ as follows:		
	\vspace{-0.4cm}
		\begin{equation}
		g_{1f}\!\left(\boldsymbol{p}_{(f)}\right)=\tr{max}\left\{0,\sum_{r \in \mathcal{R}}p_{fr}- \bar{P}_f\right\}^{2},
		\end{equation}
		and
		\vspace{-0.4cm}
		\begin{equation}
		g_{2f}\!\left(\boldsymbol{P}\right)=\tr{max}\left\{0,\psi_{f}\!\left(\boldsymbol{P}, \boldsymbol{a}_{f},\boldsymbol{S},\boldsymbol{W}\right)-\bar{C_{f}}\right\}^{2}.
		\end{equation}
		It is worth mentioning that $g_{2f}$ is a function of the other variables as well, i.e., $\boldsymbol{S},\boldsymbol{W}$ and $\boldsymbol{a}_{f}$, which are kept fixed during this power allocation step.
		

Next, we express the augmented Lagrangian function of problem~\eqref{eq:powerAllocationProblem}, defined as the regular Lagrangian function augmented by a quadratic function of the constraints, as follows:
\small
\begin{align}\label{eq:augmentedLagrangianPower}
\mathcal{L}_{1}\left(\boldsymbol{P}, \boldsymbol{\mu}_{1},\boldsymbol{\mu}_{2}\right)=& - \Theta(\boldsymbol{P}, \boldsymbol{A},\boldsymbol{S},\boldsymbol{W})+\frac{\delta}{2}\norm{\boldsymbol{g}_{1}\!\!\left(\boldsymbol{P}\right)}_{2}^{2}+\frac{\delta}{2}\norm{\boldsymbol{g}_{2}\!\left(\boldsymbol{P}\right)}_{2}^{2}+\sum_{f=1}^{F}\mu_{1f}g_{1f}\!\!\left(\boldsymbol{p}_{(f)}\right)+\sum_{f=1}^{F}\mu_{2f}g_{2f}\!\left(\boldsymbol{P}\right),
\end{align}
\normalsize
where $\boldsymbol{g}_{1}(\boldsymbol{P})=\left[g_{11}\left(\boldsymbol{p}_{(1)}\right),\ldots,g_{1f}\left(\boldsymbol{p}_{(f)}\right)\right]^{\text{T}}$, $\boldsymbol{g}_{2}(\boldsymbol{P})=\left[g_{21}(\boldsymbol{P}),\ldots,g_{2f}(\boldsymbol{P})\right]^{\text{T}}$. $\boldsymbol{\mu}_{1} \in \mathbb{R}^{F}$ and $\boldsymbol{\mu}_{2}\in \mathbb{R}^{F}$ are the Lagrangian multipliers that correspond to \eqref{eq:constraint1} and \eqref{eq:constraint2}, respectively.
$\delta$ is a positive constant and $\norm{.}$  denotes the Euclidean norm.

The ADMM consists of alternating between two steps; see~\cite{Wang2017} and references therein. The first step consists of optimizing the augmented Lagrangian~\eqref{eq:augmentedLagrangianPower} for each direction of the primal variables, while the variables for other directions remain fixed. More specifically, in our case, $\boldsymbol{p}_{r}$ is optimized for every $r \in \mathcal{R}$ by minimizing the augmented Lagrangian function. Let $\boldsymbol{p}_{r}^{(k)}$ be the value of variable $\boldsymbol{p}_{r}$ at iteration $k$. For completeness, we next describe the underlying optimization steps proposed by ADMM in the context of our problem.

\begin{proposition}
At the $k$-th iteration of the power allocation algorithm, first, the primal variables are updated as follows, for $r=1$ to $R$:
\begin{equation}\label{eq:power_allocation_solution}
\boldsymbol{p}_{r}^{(k)}=\underset{\boldsymbol{p}_{r}}{\tr{argmin}}\ \mathcal{L}_{1}\left(\boldsymbol{P}_{\bar{r}}^{(k)},\boldsymbol{\mu}_{1}^{(k-1)},\boldsymbol{\mu}_{2}^{(k-1)}\right),
\end{equation}
where
$\boldsymbol{P}_{\bar{r}}^{(k)}=\left[\boldsymbol{p}_{1}^{(k)};\ldots;\boldsymbol{p}_{r-1}^{(k)}; \boldsymbol{p}_{r};\boldsymbol{p}_{r+1}^{(k-1)}\ldots;\boldsymbol{p}_{r}^{(k-1)}\right].$
Second, the dual variables (i.e., the Lagrangian multipliers) are updated as follows:
\begin{equation}
\boldsymbol{\mu}_{1}^{(k)}=\boldsymbol{\mu}_{1}^{(k-1)}+\delta \boldsymbol{g}_{1}\!\left(\boldsymbol{P}^{(k)}\right)\text{ and }
\boldsymbol{\mu}_{2}^{(k)}=\boldsymbol{\mu}_{2}^{(k-1)}+\delta\boldsymbol{g}_{2}\!\left(\boldsymbol{P}^{(k)}\right).
\end{equation}
				The algorithm stops at convergence, once all constraints are satisfied.
\end{proposition}

We finally note that this paper uses the interior-point method \cite{Boyd2004} to solve problem~\eqref{eq:power_allocation_solution}. The simulations results of the paper particularly show the appreciable performance improvement of the adopted algorithm.

\section{Proposed NOMA Power Splitting Method}\label{sec:PowerSplitOptimization}
This section details the last step of the algorithm  which consists of optimally dividing the power allocated to every RB to its strong and weak users. The optimization problem~\eqref{eq:Optimization}  now boils down to:
\vspace{-0.7cm}
\small
\begin{subequations}\label{eq:PowerSplit}
	\begin{align}
	\underset{\substack{\boldsymbol{A}} }{\tr{max}}&\ \Theta(\boldsymbol{P}, \boldsymbol{A},\boldsymbol{S},\boldsymbol{W})\\
				\tr{s.t.\ \ } 	& 0 \leq a_{fr}\leq 1,\ \forall (f,r)\in \left(\mathcal{F}\times\mathcal{R}\right),\\
		&\psi_{f}\!\left(\boldsymbol{P}, \boldsymbol{a}_{f},\boldsymbol{S},\boldsymbol{W}\right)\leq \bar{C_{f}},\  \forall f\in \mathcal{F}
	\end{align}
\end{subequations}
\normalsize
This optimization is separable per FAP, i.e., solving~\eqref{eq:PowerSplit} is equivalent to solving the following optimization for every FAP:
				\begin{subequations}\label{eq:powersplitFAP}
				\begin{align}
				\underset{\substack{\boldsymbol{a}_{f}} }{\tr{max}}&\   \Theta_{f}(\boldsymbol{P}, \boldsymbol{a}_{f},\boldsymbol{S},\boldsymbol{W})=\sum_{r\in \mathcal{R}}\sum_{u\in \mathcal{U} }s_{fru}\alpha_{u}C^{(s)}_{fru}(\boldsymbol{p}_{r},a_{fr})+w_{fru}\alpha_{u}C^{(w)}_{fru}(\boldsymbol{p}_{r},a_{fr})\label{eq:newobjectivePS}\\
				\tr{s.t.\ \ } 	& 0 \leq a_{fr}\leq 1,\ \forall r\in \ \mathcal{R},\\
				&\psi_{f}\!\left(\boldsymbol{P}, \boldsymbol{a}_{f},\boldsymbol{S},\boldsymbol{W}\right)\leq \bar{C_{f}}.
				\end{align}
				\end{subequations}
Therefore, this step of the algorithm can be solved at every FAP in order to alleviate the traffic on the fronthaul link in alignment to the FRAN principle of pushing some control to the network edge.
Hence, problem~\eqref{eq:powersplitFAP} can be re-written as follows:
\small
				\begin{subequations}\label{eq:PS_optimization2}
				\begin{align}
				\underset{\substack{\boldsymbol{a}_{f} }}{\tr{min}}&\ - \Theta_{f}\!\left(\boldsymbol{P}, \boldsymbol{a}_{f},\boldsymbol{S},\boldsymbol{W}\right)\\
				\tr{s.t.\ \ } 	& -\boldsymbol{a}_{f}\leq\boldsymbol{0}\\
										& \boldsymbol{a}_{f}-\boldsymbol{1} \leq\boldsymbol{0}\\
				&\psi_{f}\!\left(\boldsymbol{P}, \boldsymbol{a}_{f},\boldsymbol{S},\boldsymbol{W}\right)-\bar{C_{f}}\leq 0.
				\end{align}
				\end{subequations}
\normalsize
Moreover, the utility function at FAP $f$ is separable per RB. More specifically, in~\eqref{eq:newobjectivePS}, the $r$-th term in the summation over the RBs is a function of $a_{fr}$ only. This means that, once again, problem~\eqref{eq:PS_optimization2} can be efficiently solved by ADMM. Towards this end, the constraints are reformulated in a similar way as for the power allocation to every RB (cf. Sec.~\ref{sec:PowerAllocation}	).  In what follows, only the variables under consideration $(\boldsymbol{a}_{f} )$  are shown in order to simplify the mathematical notations. The resulting optimization which is solved at the FAP $f$ is
\small
				\begin{subequations}\label{eq:PS_feformulation}
				\begin{align}
				\underset{\substack{\boldsymbol{a}_{f} }}{\tr{min}}&\  - \Theta_{f}\!\left(\boldsymbol{a}_{f}\right)\\
				\tr{s.t.\ \ } 	& \tr{max}\left\{0,-a_{fr}\right\}^{2}=0 ,\ \forall r\in \ \mathcal{R},\label{eq:constr1}\\
										&\tr{max}\left\{0,a_{fr}-1\right\}^{2} =0,\  \forall r\in \ \mathcal{R},\label{eq:constr2}\\
				&\tr{max}\left\{0,\psi_{f}\!\left(\boldsymbol{a}_{f}\right)-\bar{C_{f}}\right\}^{2}=   0\label{eq:constr3}.
				\end{align}
				\end{subequations}	
\normalsize				
The augmented Lagrangian function is now:
\small
\begin{align}\label{eq:LagrangianPS}
				\mathcal{L}_{2}\left(\boldsymbol{a}_{f},\boldsymbol{\lambda}_{1},\boldsymbol{\lambda}_{2},\lambda_{3}\right)=& - \Theta_{f}(\boldsymbol{a}_{f})+\frac{\delta}{2}\sum_{r=1}^{R}\norm{\tr{max}\left\{0,-a_{fr}\right\}^{2}}_{2}^{2}+\sum_{r=1}^{R}\lambda_{1r}\tr{max}\left\{0,-a_{fr}\right\}^{2}+\nonumber\\
				&\frac{\delta}{2}\sum_{r=1}^{R}\norm{\tr{max}\left\{0,a_{fr}-1\right\}^{2}}_{2} ^{2}+\sum_{r=1}^{R}\lambda_{2r}\tr{max}\left\{0,a_{fr}-1\right\}^{2}+\nonumber\\
				&\frac{\delta}{2}\norm{\tr{max}\left\{0,\psi_{f}\!\left(\boldsymbol{a}_{f}\right)-\bar{C_{f}}\right\}^{2}}_{2}^{2}+\lambda_{3}\tr{max}\left\{0,\psi_{f}\!\left(\boldsymbol{a}_{f}\right)-\bar{C_{f}}\right\}^{2},
				\end{align}
\normalsize				
\noindent where $\boldsymbol{\lambda}_{1} \in \mathbb{R}^{F}$, $\boldsymbol{\lambda}_{2}\in \mathbb{R}^{F}$ and $\lambda_3 \in \mathbb{R}$ are the Lagrangian multipliers that correspond to \eqref{eq:constr1}, \eqref{eq:constr2} and \eqref{eq:constr3}, respectively, and where $\delta$ is a positive constant.
The first step of every ADMM iteration consists of minimizing the Lagrangian function~\eqref{eq:LagrangianPS} for every direction $r$, while the other variables corresponding to all other directions are fixed. The second step consists of updating the dual variables ~\cite{Wang2017}.
\begin{proposition}
At the $k$-th iteration of the NOMA power splitting optimization algorithm at FAP $f$, first, the primal variables are updated as follows, for $r=1$ to $R$:
\begin{align}\label{eq:NOMALagrangianMin}
				&{a}_{fr}^{(k)}=\underset{{a}_{fr}}{\tr{argmin}}\ \mathcal{L}_{2}\left(\boldsymbol{a}_{f\bar{r}},\boldsymbol{\lambda}_{1},\boldsymbol{\lambda}_{2},\lambda_{3}\right).
				\end{align}	
				where $\boldsymbol{a}_{f\bar{r}}=\left[a_{f1}^{(k)},\ldots,a_{f(r-1)}^{(k)}, a_{fr},a_{f(r+1)}^{(k-1)},\ldots, a_{fr}^{(k-1)}\right]$.
Second,  the dual variables are updated as follows:
\small
\begin{subequations}
				\begin{align}
				\lambda_{1r}^{(k)}&=\lambda_{1r}^{(k-1)}+\rho\tr{max}\left\{0,-a_{fr}^{(k)}\right\}^{2},\\
				\lambda_{2r}^{(k)}&=\lambda_{2r}^{(k-1)}+\rho\tr{max}\left\{0,a_{fr}^{(k)}-1\right\}^{2},\\
				\lambda_{3}^{(k)}&=\lambda_{3}^{(k-1)}+\rho\tr{max}\left\{0,\psi_{f}\!\left(\boldsymbol{a}_{f}^{(k)}\right)-\bar{C_{f}}\right\}^{2}.
				\end{align}
				\end{subequations}
				\normalsize
				The algorithm stops at convergence, once all constraints are satisfied.
\end{proposition}
Once again, this paper utilizes the interior-point method to solve problem~\eqref{eq:NOMALagrangianMin} as well.



\section{Complexity Analysis} \label{sec:ComplexityAnalysis}	
This section describes the computational complexity of each algorithm step described in the previous sections.
Reference \cite{Edmonds1972} shows that the complexity of the Hungarian method is in the order of $O(n^{3})$, where $n$ is the number of agents. For the Hungarian-based assignment adopted in our paper, since the assignment is done for every RB, the agents are equal to the number of users that can be assigned, i.e., $2F$. The complexity of the MCKP-greedy algorithm is $O\left(\sum_{i=1}^{m}n_{i}\log(n_{i})+n\log(n)\right)$ \cite{Kellerer2004}, where $m$ is the number of classes of objects to be put in one knapsack and $n_{i}$ is the number of elements in the $i$-th class. In our case, $m=R$ and $n_{i}={U\choose 2},\  \forall i \in \{1,\ldots, R\}$. Lastly, $n$ is the total number of elements for all classes, i.e., $n=R{U\choose 2} $. Each main iteration of the interior-point method  has a worst-case complexity which is a polynomial function of the problem size \cite{Quan2006}. More specifically, the total number of arithmetic operations needed to perform the interior-point method is in the order of $O(n^{3})$, where $n$ is the number of variables.
Based on the above, Table~\ref{table:algoComplexity} displays the complexity of each algorithm and the corresponding number of repetitions required to implement the overall algorithm~\ref{alg:overallAlgorithm}.
		\renewcommand{\arraystretch}{1.3}
\begin{table}[h!]
\centering
 \begin{tabular}{c c c}
 \hline
\textbf{Algorithm} & \textbf{Complexity} & \textbf{Repetitions}  \\ [0.5ex]
 \hline\hline
Hungarian algorithm& $O((2F)^{3}) $& $R$  \\
 \hline
 MCKP-greedy algorithm  & $O\left(Ru\log\left(Ru^{2}\right)\right), u={U\choose 2} $& $FA$\\
 \hline
Power allocation  &  $O(F^{3})$ & $R$ \\
 \hline
Power split optimization  &  $O(1^{3})$ & $FR$ \\
 \hline
\end{tabular}
\caption{Complexity of each algorithm and the number of repetitions needed for implementing the 
user-to-RB assignment, power allocation and NOMA power splitting }
\label{table:algoComplexity}
\end{table}
The number of repetitions for  MCKP-greedy algorithm is $RA$, where $A$ is the total number of auction iterations.

Table~\ref{table:algoComplexity} shows that taking advantage of the problem structure in order to apply ADMM to solve both continuous parts of the problem not only allows the distribution of the algorithm tasks to all FAPs, but also significantly reduces the computational complexity. Thanks to ADMM, the complexity of the power allocation is reduced from $O((FR)^{3})$ to $O(RF^{3})$. For the power split optimization, it is reduced from $O((FR)^{3})$ to $O(FR)$.


\section{Simulation Results}\label{sec:SimulationResults}

\subsection{Simulation Setup}\label{subsec:simulation_setup}
This section evaluates the performance of the proposed algorithms in a grid network of 7 FAPs with wrap-around architecture\cite{Panwar2017}. Four users are uniformly distributed inside each FAP's coverage.
All simulation results are averaged over 10000 channel realizations.
We assume that each FAP has a bandwidth of $W=10$ MHz equally divided among its RBs, i.e., $\beta=\frac{W}{R}$.
The 3GPP urban micro-cell environment and simulation parameters are adopted as in~\cite{3gpp}. The inter-FAP distance is fixed to $200m$. The path-loss models both the signal power attenuation and the channel shadowing effect, and is given by  $l(d,f_c)=36.7\log{10}(d)+22.8+20\log{10}(f_c)$, where $d$ is the distance in meters, and $f_c$, the carrier frequency, is set to $f_c$=2.5 GHz. All channels undergo Rayleigh fading. We assume that the available power and the fronthaul capacity limits are equal for all FAPs. While the power is set to $\bar{p}_{f}=41\text{dBm}$ $\forall f\in \mathcal{F}$, $\bar{C}_f=\bar{C}$ changes across the simulations so as to illustrate the impact of backhaul capacity on the proposed algorithms performance. Unless specified otherwise, the number of RBs is set to $R=2$, and the averaging time window $\tau$ is set to $\tau=50$ (see more info about $\tau$ below). We assume perfect SIC, i.e., $\zeta=0$.

For illustrative purposes, the user weights are taken for two extreme cases:
\begin{enumerate}
\item  $\alpha_u=1$ for all users, i.e., sum-rate maximization.
\item  $\alpha_{u}=\frac{1}{\bar{C}_{u}^{(\tau)}}$, the inverse of the average user rate $\bar{C}_{u}^{(\tau)}$  over a time window $\tau$, i.e., proportional fair weighted sum-rate. At initialization, $\bar{C}_u$ is fixed assuming that user $u$ is assigned to its closest FAP, and all channel gains are equal to 1.
\end{enumerate}

In addition to the utility values and rates, the algorithms' performance is also evaluated in terms of Jain's fairness index between users, defined as:
			
			\begin{equation}
			\mathcal{J}=\left. \left(\displaystyle\sum_{u=1}^{U}C_{u}\right)^{2}\middle/ U\displaystyle\sum_{u=1}^{U}C_{u}^{2}\right. .
			\end{equation}
			\normalsize
			
			\noindent Its value ranges from $1/U$ (worst case) to 1 (best case), which is achieved when all users are served with equal rates.

\subsection{Baseline Schemes and Acronyms}
We now define the following baseline schemes for bench-marking purposes.
	\begin{enumerate}
	\item Voronoi-based assignment: the baseline scheme for proposed user assignment, where two users inside every FAP's  Voronoi cell are randomly assigned to each of its RBs. The user with lower channel gain is the weak user, while the user with higher channel gain is the strong user.
		\item Uniform power allocation: the baseline scheme for proposed power optimization. Within every FAP if Hungarian  or Voronoi assignment is applied, it is not guaranteed that the resulting assignment satisfies the fronthaul capacity limitation. Therefore, for the FAPs with non-satisfied fronthaul capacity constraints, the RBs' powers are uniformly decreased until they are satisfied.
	\item OMA: this is the conventional allocation scheme used in current systems such as OFDMA in LTE-Advanced, where each basic resource unit (subchannel or
RB of bandwidth $\beta$) is allocated to only one user.
	\end{enumerate}		

	\renewcommand{\arraystretch}{1}			
		\begin{table}[h!]
				\centering
				\small
				\begin{tabular}{ c  c ||  c    c || c    c }
				\hline
					V & Voronoi-based assignment &	PU&  Uniform power allocation& FPS&  Fixed NOMA power split\\ [0.5ex]
					\hline
					H & Hungarian-based assignment & PA& Optimized power allocation & PS& Optimized power split  \\ [0.5ex]
					\hline\hline
					K & MCPK-based assignment &  & \\
					\hline
				
				\end{tabular}
				\caption{Summary of the notations in the figures.}
				\label{table:Figure_notations}
	\end{table}	
	
	For sake of clarity, Table~\ref{table:Figure_notations} summarizes the acronyms used in the figures.

\subsection{Effect of initialization of $\boldsymbol{A}$}
\begin{figure}[t!]
    \centering
    \begin{subfigure}[t]{0.47\textwidth}
      \psfrag{aooooo 005}[c][c][0.5]{ $\boldsymbol{A}^{(o)}=0.05$}
          \psfrag{aooooo 025}[c][c][0.5]{ $\boldsymbol{A}^{(o)}=0.25$}
        \psfrag{aooooo 05}[c][c][0.5]{ $\boldsymbol{A}^{(o)}=0.5$}
        \psfrag{aooooo 075}[c][c][0.5]{ $\boldsymbol{a}_o=0.75$}
        \psfrag{aooooo 095}[c][c][0.5]{ $\boldsymbol{A}^{(o)}=0.95$}
      \psfrag{CDF}[c][c][0.6]{Utility CDF}
      \psfrag{Utility Value}[c][c][0.6]{Utility value, WSR.}
        \includegraphics[width=\textwidth]{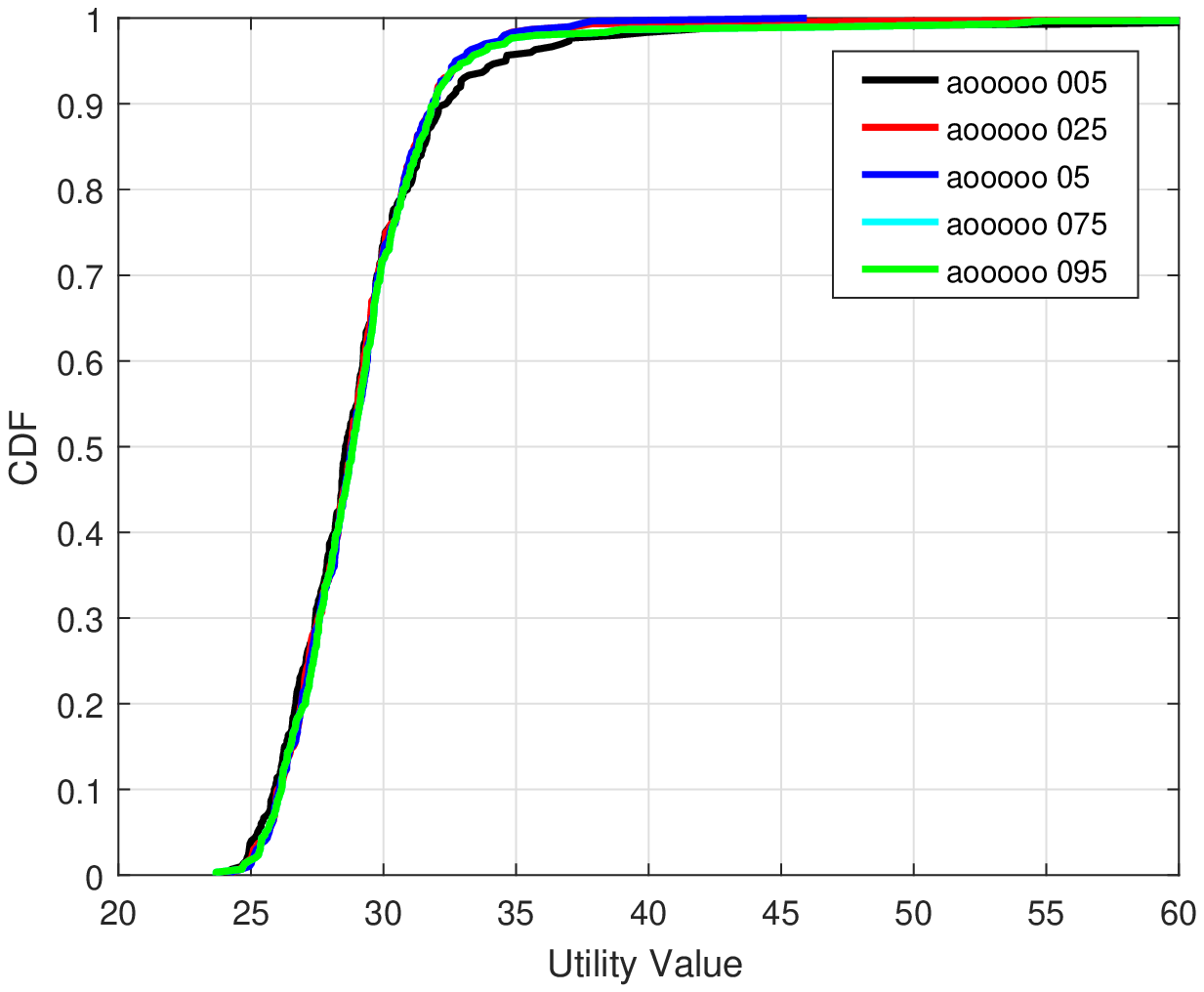}
        \caption{\small Weighted sum-rate maximization}
        \label{fig:Utility_a_o_wsr}
    \end{subfigure}\hspace{-0.0\textwidth}
    \begin{subfigure}[t]{0.47\textwidth}
      \psfrag{aooooo 005}[c][c][0.5]{ $\boldsymbol{A}^{(o)}=0.05$}
          \psfrag{aooooo 025}[c][c][0.5]{ $\boldsymbol{A}^{(o)}=0.25$}
        \psfrag{aooooo 05}[c][c][0.5]{ $\boldsymbol{A}^{(o)}=0.5$}
        \psfrag{aooooo 075}[c][c][0.5]{ $\boldsymbol{A}^{(o)}=0.75$}
        \psfrag{aooooo 095}[c][c][0.5]{ $\boldsymbol{A}^{(o)}=0.95$}
              \psfrag{CDF}[c][c][0.6]{Utility CDF}
         \psfrag{Utility Value, SR}[c][c][0.6]{Utility value, SR.}
        \includegraphics[width=\textwidth]{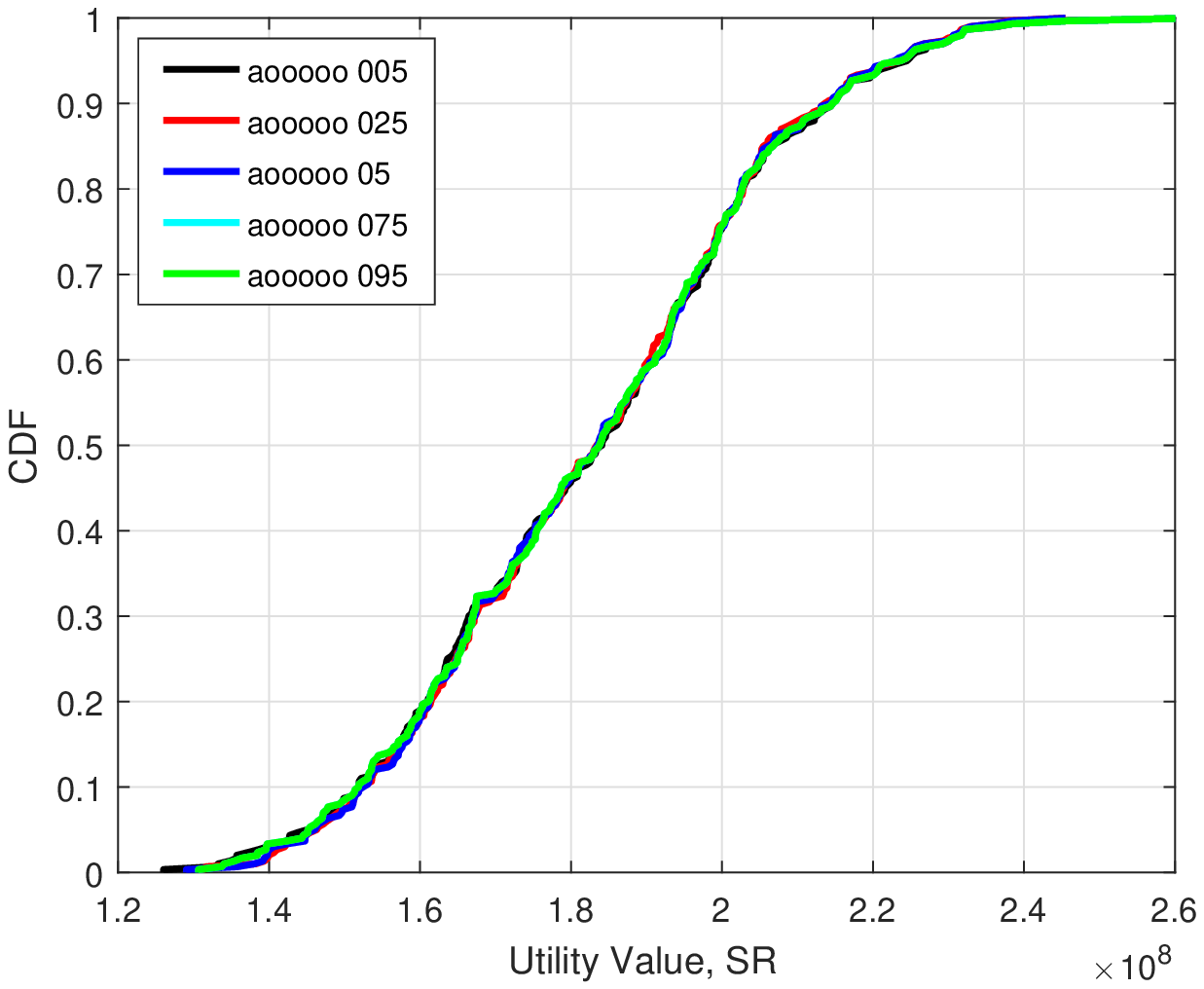}
         \caption{\small Sum-rate maximization}
         \label{fig:Utility_a_o_sr}
 \end{subfigure}
    \caption{CDF of the utility function for different values of $\boldsymbol{A}^{(o)}$, $\bar{C}_{f}=10^{8}$. }
    \label{fig:Utility_a_o}
\end{figure}

First, we simulate the effect of the NOMA power splitting initialization matrix $\boldsymbol{A}^{(o)}$ on different network performance metrics.
Fig.~\ref{fig:Utility_a_o} shows the CDF of the utility function for different values of $\boldsymbol{A}^{(o)}$ for H-PA-PS, for weighted sum-rate maximization in Fig.~\ref{fig:Utility_a_o_wsr}  and for sum-rate maximization in Fig.~\ref{fig:Utility_a_o_sr}. The corresponding Jain's fairness index is shown in Table~\ref{table:Jains_a_o}.

	\renewcommand{\arraystretch}{1.3}			
		\begin{table}[h!]
				\centering
				\footnotesize
				\begin{tabular}{c c c c c c}
					\hline
					{$\boldsymbol{A}^{(o)}$}&$0.05$& $0.25$& $0.5$&$0.75$&$0.95$\\ [0.5ex]
					\hline\hline
					WSR & 0.40 & 0.41& 0.40& 0.38& 0.37 \\
					\hline
							SR& 0.23 & 0.22& 0.23& 0.23& 0.24 \\
					\hline
				\end{tabular}
				\caption{Jain's fairness indexes for different $\boldsymbol{A}^{(o)}$.}
				\label{table:Jains_a_o}
	\end{table}

			It can be seen that, in general, the value of $\boldsymbol{A}^{(o)}$ does not significantly affect the behavior of both utility functions. Table~\ref{table:Jains_a_o}, however, shows that under weighted sum-rate maximization, the best user fairness is achieved at $\boldsymbol{A}^{(o)}$ around 0.25. For sum-rate maximization, the fairness is not affected by the value of $\boldsymbol{A}^{(o)}$.

Therefore, in all subsequent results, we set $\boldsymbol{A}^{(o)}=0.25$ for the proposed NOMA-based algorithms.

\subsection{Distribution of $\boldsymbol{a}$ }
\begin{figure}[t!]
    \centering
    \psfrag{cdf}[c][c][1]{CDF of $\boldsymbol{a}$ }
    \psfrag{a}[c][c][1]{ $\boldsymbol{a}$ }
      \psfrag{SR----}[c][c][0.9]{ SR }
      \psfrag{WSR 50----}[c][c][0.9]{ $\quad$WSR $\tau=50$ }
      \psfrag{WSR 25----}[c][c][0.9]{$\quad$ WSR $\tau=25$ }
       \psfrag{Hungarian--}[c][c][0.9]{ H-PA-PS}
        \psfrag{Knapsack--}[c][c][0.9]{ K-PA-PS}
       \scalebox{0.5}{ \includegraphics[width=\textwidth]{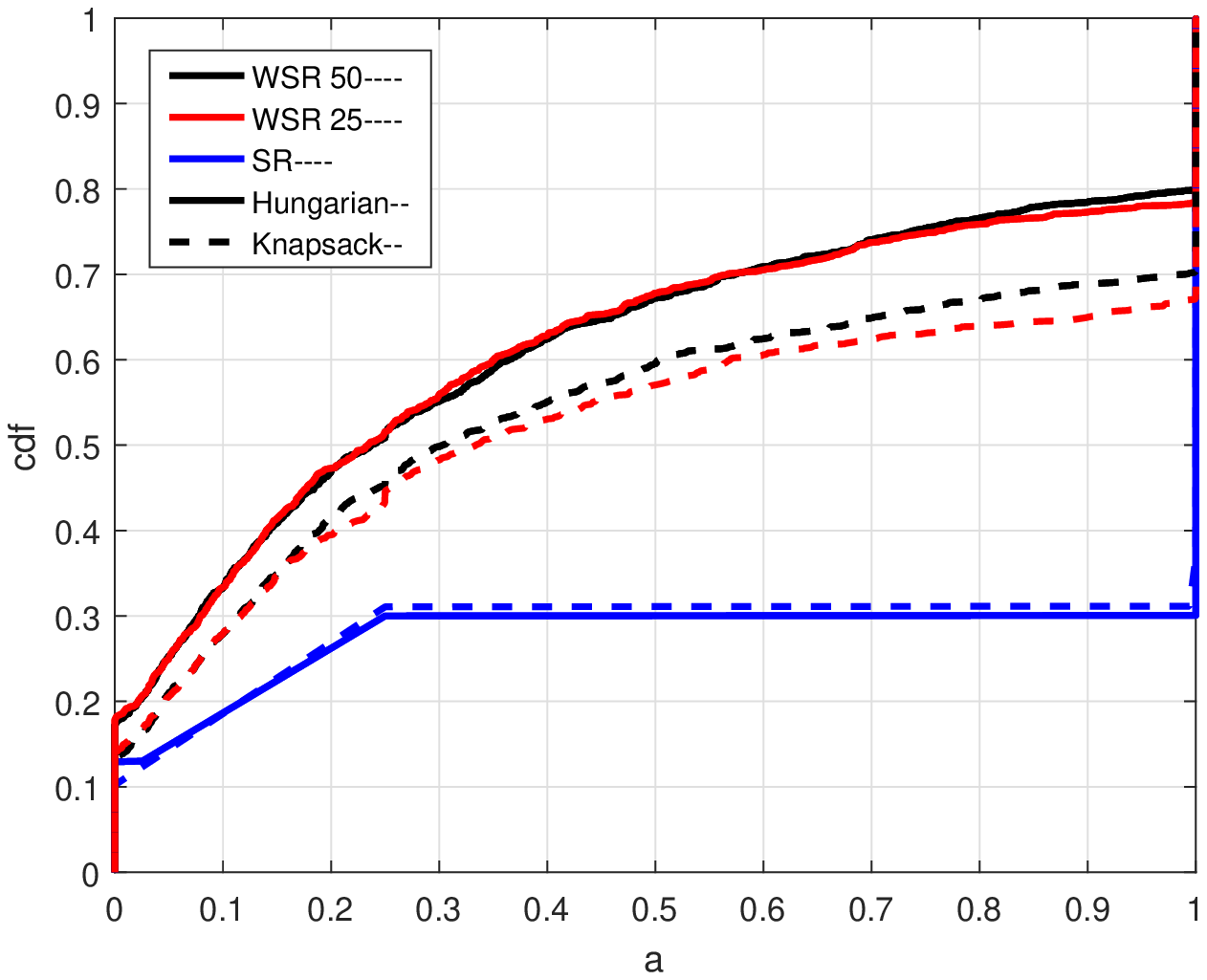}}
        \caption{CDF of the value of $\boldsymbol{a}$ after power split optimization.}
        \label{fig:aDistribution}
\end{figure}

Fig.~\ref{fig:aDistribution}  shows the distribution of the  NOMA power split ratios after optimization, i.e., the power distribution among weak and strong users. It shows that, for sum-rate maximization and for both H-PA-PS and  K-PA-PS, full power is given to the strong user for 70\% of the cases.
For weighted sum-rate maximization, the window size $\tau$ does not significantly change the distribution of the NOMA power split ratios.
The probability that the strong user is served with full power for K-PA-PS is about 30\%, while it is 20\% for H-PA-PS. The weak user gets full power for 15\% and 18\% of the time for  K-PA-PS and  H-PA-PS, respectively.
For both algorithms, $\boldsymbol{a}$ smoothly varies between 0 and 0.5 for about 50\% of the cases under weighted sum-rate maximization. This figure, therefore, shows that the proposed algorithms, under weighted sum-rate maximization, indeed instigate fairer power allocation compared to sum-rate maximization case.

\subsection{Algorithm Performance}\label{subsec:algo_performance}
\begin{figure}[t!]
    \centering
    \begin{subfigure}[t]{0.47\textwidth}
    \psfrag{Number of iterations}[c][c][0.6]{ Number of iterations of the main loop  of Algorithm \ref{alg:overallAlgorithm}  }
    \psfrag{K-PA-PS---}[c][c][0.5]{K-PA-PS }
     \psfrag{H-PA-PS---}[c][c][0.5]{H-PA-PS }
      \psfrag{K-PS---}[c][c][0.5]{$\quad$K-PU-PS}
      \psfrag{K---}[c][c][0.5]{$\quad \quad \quad $K-PU-FPS}
      \psfrag{H-PA---}[c][c][0.5]{$\quad \ $H-PA-FPS}
       \psfrag{V-PU---}[c][c][0.5]{$\quad \ $V-PU-FPS}
        \psfrag{Utility function value}[c][c][0.6]{Utility function (weighted sum-rate)}
        \includegraphics[width=\textwidth]{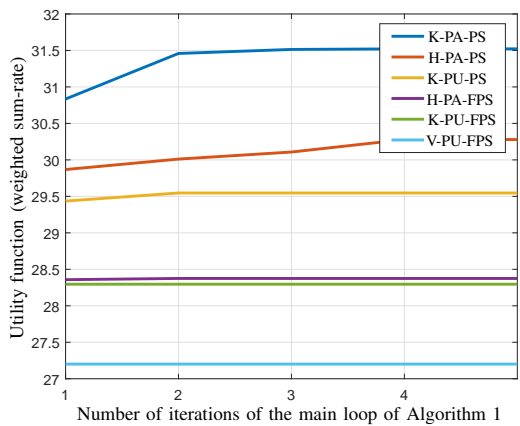}
        \caption{\small Weighted sum-rate maximization, $\tau=50$}
        \label{fig:convergence_wsr}
    \end{subfigure}\hspace{-0.0\textwidth}
    ~
    \begin{subfigure}[t]{0.47\textwidth}
        \psfrag{Number of iterations}[c][c][0.6]{ Number of iterations of the main loop  of Algorithm \ref{alg:overallAlgorithm}  }
    \psfrag{K-PA-PS---}[c][c][0.5]{K-PA-PS }
     \psfrag{H-PA-PS---}[c][c][0.5]{H-PA-PS }
      \psfrag{K-PS---}[c][c][0.5]{$\quad$K-PU-PS}
      \psfrag{K}[c][c][0.5]{$\quad \quad \quad \  $K-PU-FPS}
      \psfrag{H-PA---}[c][c][0.5]{$\quad \ $H-PA-FPS}
       \psfrag{V-PU---}[c][c][0.5]{$\quad \ $V-PU-FPS}
        \psfrag{Utility function value}[c][c][0.6]{Utility function}
        \includegraphics[width=\textwidth]{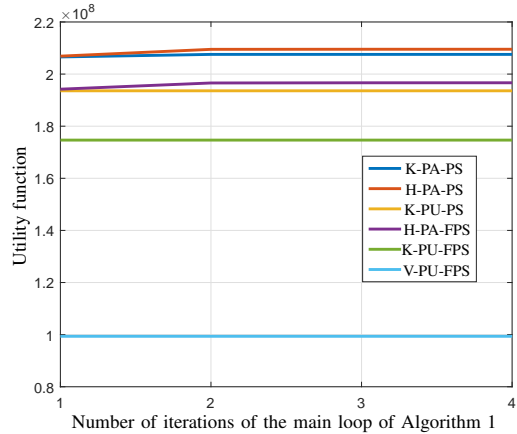}
         \caption{\small Sum-rate maximization}
         \label{fig:convergence_sr}
 \end{subfigure}
    \caption{Convergence of different algorithms, $\bar{C}_{f}=10^{8}$. }
    \label{fig:convergence}
\end{figure}

Fig.~\ref{fig:convergence} shows the performance for each utility function according to the outside loop number of iterations in algorithm~\ref{alg:overallAlgorithm}, for weighted sum-rate (Fig.~\ref{fig:convergence_wsr}) and sum-rate maximization (Fig.~\ref{fig:convergence_sr}). The performance for each utility function is shown for different combinations of algorithms inside the loop.
The figure shows that for all combinations, if more than one iteration is needed, the algorithm converges within a few number of iterations. As for the inner loops, for the power allocation and the NOMA power splitting, ADMM converges within less than 5 iterations.

		
Under both utility functions, the figures show that, the proposed algorithm (partially or wholly applied) outperforms the baseline scheme (Voronoi assignment combined with uniform power allocation).
In more details, for weighted sum-rate, the proposed K-PA-PS yields to the best network utility value, with a 16\%-gain compared to baseline V-PU-FPS, while proposed H-PA-PS yields to 11\%-gain compared to V-PU-FPS.
H-PA-PS is, however, slightly better than K-PA-PS for sum-rate maximization (111\% vs 109\% for K-PA-PS).

As previously mentioned, the proposed MCKP assignment algorithm can be applied alone without optimized power allocation, as its result is always feasible in terms of fronthaul capacity limitation. In this case, each FAP's power budget is equally divided between its RBs and the power split is fixed to 0.25 (noted K) or optimized (noted K-PS).
It is observed that K-PU-FPS performs 4\% and 76\% better than V-PU-FPS for weighted sum-rate and sum-rate, respectively. Combined with optimized power split, K-PU-PS yields to 8.6\% and 95\% gain, for weighted sum-rate and sum-rate, respectively. Combined with optimized power allocation, the Hungarian assignment H-PA-PS yields to 4.3\% and 98\% gain compared to V-PU-FPS for weighted sum-rate and sum-rate, respectively.

\begin{figure}[t!]
    \centering
    \begin{subfigure}[t]{0.47\textwidth}
    \psfrag{K-PA-PS---, Cr=10 8}[c][c][0.5]{$\quad$K-PA-PS, $C_f=10^{8}$}
    \psfrag{H-PA-PS---, Cr=10 8}[c][c][0.5]{$\quad$H-PA-PS, $C_f=10^{8}$}
    \psfrag{K-PA-PS-SR---, Cr=5*10 7}[c][c][0.5]{K-PA-PS, $C_f=5\times 10^{7}$}
     \psfrag{H-PA-PS---, Cr=5*10 7}[c][c][0.5]{$\quad$H-PA-PS, $C_f=5\times 10^{7}$}
     \psfrag{Utility Value}[c][c][0.6]{ Utility value, weighted sum-rate.}
     \psfrag{CDF}[c][c][0.6]{CDF}
        \includegraphics[width=\textwidth]{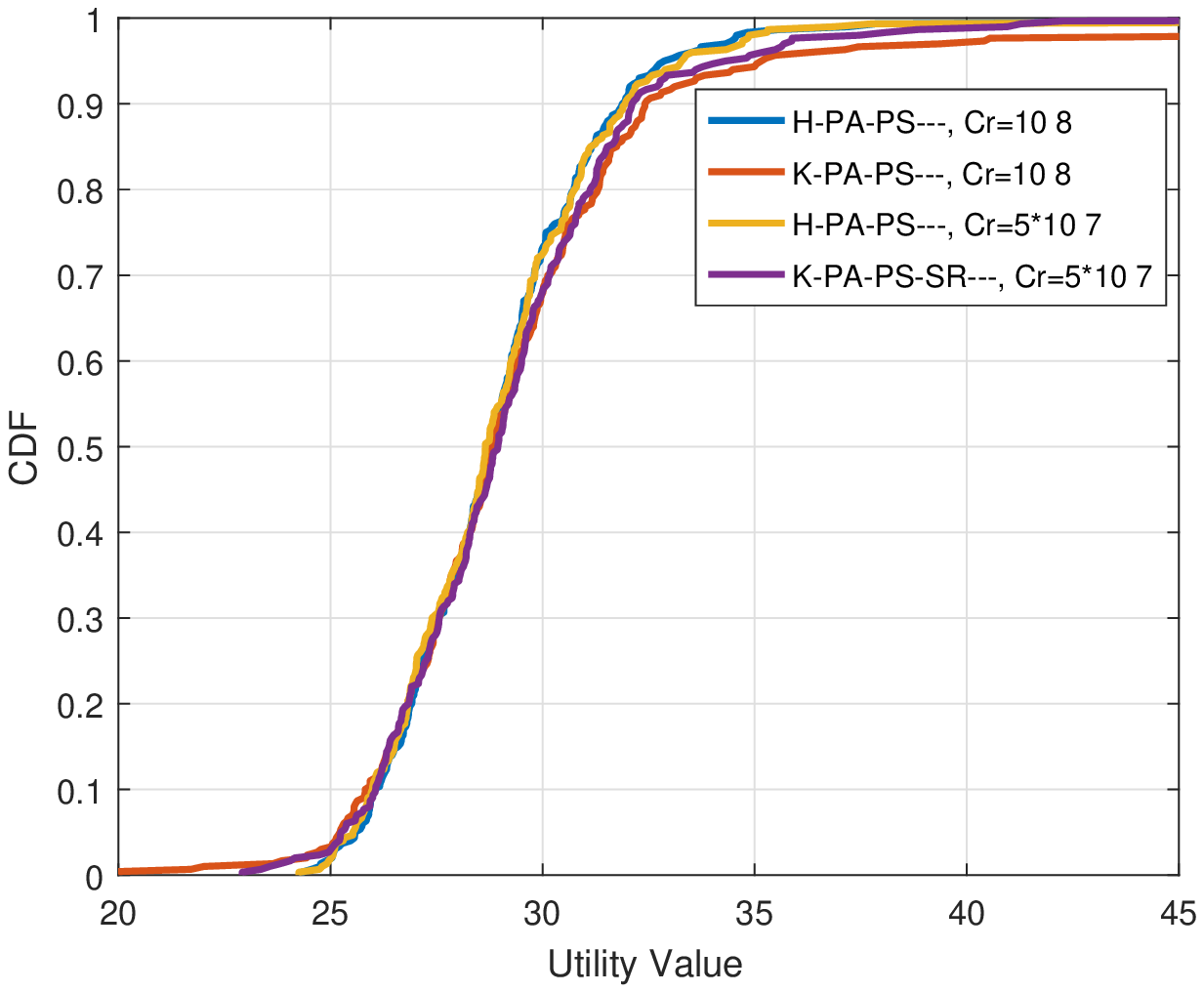}
        \caption{\small weighted sum-rate maximization.}
        \label{fig:Algorithm_wsr}
    \end{subfigure}\hspace{-0.0\textwidth}
    ~
    \begin{subfigure}[t]{0.47\textwidth}
    \psfrag{K-PA-PS---, Cr=10 8}[c][c][0.5]{$\quad$K-PA-PS, $C_f=10^{8}$}
    \psfrag{H-PA-PS---, Cr=10 8}[c][c][0.5]{$\quad$H-PA-PS, $C_f=10^{8}$}
    \psfrag{K-PA-PS-SR---, Cr=5*10 7}[c][c][0.5]{K-PA-PS, $C_f=5\times 10^{7}$}
     \psfrag{H-PA-PS---, Cr=5*10 7}[c][c][0.5]{$\quad \ $H-PA-PS, $C_f=5\times 10^{7}$}
     \psfrag{Utility Value, SR}[c][c][0.6]{Utility value,sum-rate. }
     \psfrag{CDF}[c][c][0.6]{CDF}
        \includegraphics[width=\textwidth]{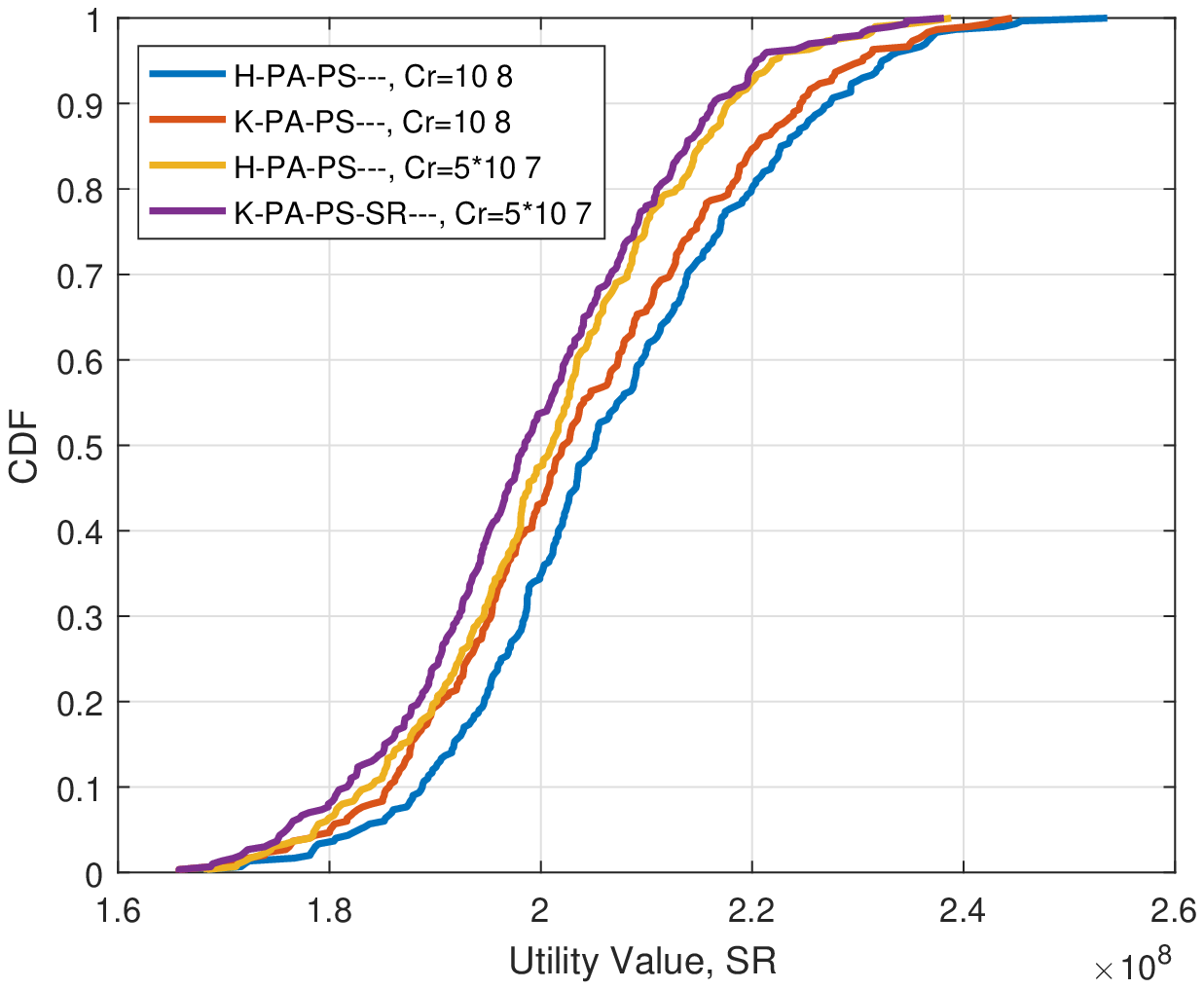}
         \caption{\small Sum-rate maximization.}
         \label{fig:Algorithm_sr}
 \end{subfigure}
    \caption{Performance comparison, for different $\bar{C}_{f}$ values. }
    \label{fig:Algorithm_comparison}
\end{figure}

Fig.~\ref{fig:Algorithm_comparison} compares the performance of H-PA-PS and K-PA-PS for different fronthaul capacity values and for weighted sum-rate (Fig.~\ref{fig:Algorithm_wsr}) and sum-rate (Fig.~\ref{fig:Algorithm_sr}) maximization. The figures confirm again that K-PA-PS is slightly better than H-PA-PS for weighted sum-rate maximization, regardless of the value $\bar{C}_f$.   H-PA-PS, however, yields to higher sum-rates. The Hungarian-based assignment is favorable to sum-rate maximization as it does not take the fronthaul capacity into consideration, i.e., it always chooses the best users for each RB.
With MCKP, the users assignment is subjected to the fronthaul capacity allowing weaker users to be served as well, thereby increasing the fairness between users.

Table~\ref{table:jain_algorithm} displays the values of the Jain's fairness indexes that correspond to the curves of Fig.~\ref{fig:Algorithm_comparison}, i.e., for  H-PA-PS and K-PA-PS for different fronthaul capacity values and for both utilities.
As expected, the Jain's fairness index is low for sum-rate maximization compared to the one of weighted sum-rate maximization.  For sum-rate maximization, its value is higher for lower values of $\bar{C}_f$.
This shows that, under sum-rate maximization, on one hand, if $\bar{C}_f$ is high enough, the network prioritizes users with good channels. On the other hand, if $\bar{C}_f$ is low, some FAPs either serve weak users in order to satisfy $\bar{C}_f$ (MCKP-based assignment) or still serve users with best channels but with lower power (Hungarian-based assignment). In either case, the fairness index increases. This phenomenon can be observed in Fig.~\ref{fig:fairnessRB} as well.


\renewcommand{\arraystretch}{1.3}
	\begin{table}[!htbp]
\centering
\begin{tabular}{|c|c|c|c|c|}
\hline
 Utility &  \multicolumn{2}{|c|}{Weighted sum-rate} & \multicolumn{2}{|c|}{Sum-rate}\\
\hline
$\bar{C}_{f}$  &$ 5 \times 10^{7}$  &   $ 10^{8}$& $ 5 \times 10^{7}$    & $ 10^{8}$ \\
\hline\hline
		 H-PA-PS  			& 0.42& 0.42& 0.25& 0.22\\
\hline
	 K-PA-PS  				& 0.41& 0.40& 0.25& 0.21\\
\hline
\end{tabular}
\caption{Jain's fairness indexes for weighted sum-rate and sum-rate maximization.}
				\label{table:jain_algorithm}\end{table}

\subsection{NOMA vs. OMA}

\begin{figure}[t!]
    \centering
    \begin{subfigure}[t]{0.48\textwidth}
        \psfrag{NOMA-------, Cr 10 8}[c][c][0.47]{NOMA, $C_f=10^{8}$}
        \psfrag{OMA-------, Cr 10 8}[c][c][0.47]{OMA, $C_f=10^{8}$}
         \psfrag{NOMA-------, Cr 5 10 7}[c][c][0.47]{$\ $NOMA, $C_f=\! 5\times 10^{7}$}
       \psfrag{OMA-------, Cr 5 10 7}[c][c][0.47]{$\quad$OMA, $C_f=5\times 10^{7}$}
  \psfrag{Utility Value, WSR}[c][c][0.6]{Utility Value, weighted sum-rate.}
  \psfrag{CDF}[c][c][0.6]{CDF}
        \includegraphics[width=\textwidth]{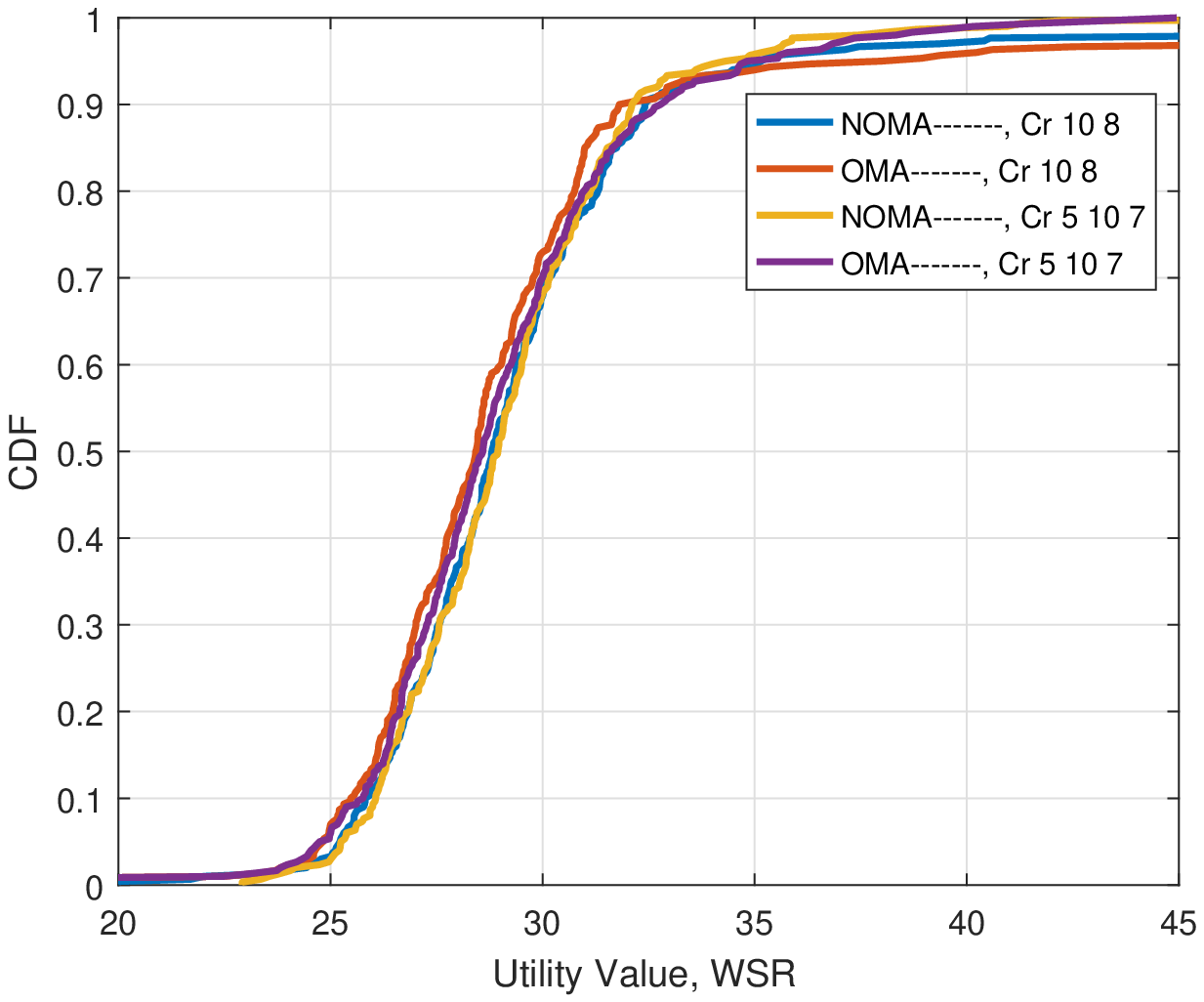}
        \caption{\small Weighted sum-rate maximization.}
        \label{fig:OMA_wsr}
    \end{subfigure}\hspace{-0.0\textwidth}
    ~
    \begin{subfigure}[t]{0.48\textwidth}
        \psfrag{NOMA-------, Cr 10 8}[c][c][0.47]{NOMA, $C_f=10^{8}$}
        \psfrag{OMA-------, Cr 10 8}[c][c][0.47]{OMA, $C_f=10^{8}$}
         \psfrag{NOMA-------, Cr 5 10 7}[c][c][0.47]{$\ $NOMA, $C_f=\! 5\times 10^{7}$}
       \psfrag{OMA-------, Cr 5 10 7}[c][c][0.47]{$\quad$OMA, $C_f=5\times 10^{7}$}
              \psfrag{Utility Value, SR}[c][c][0.6]{Utility Value, sum-rate.}
              \psfrag{CDF}[c][c][0.6]{CDF}
        \includegraphics[width=\textwidth]{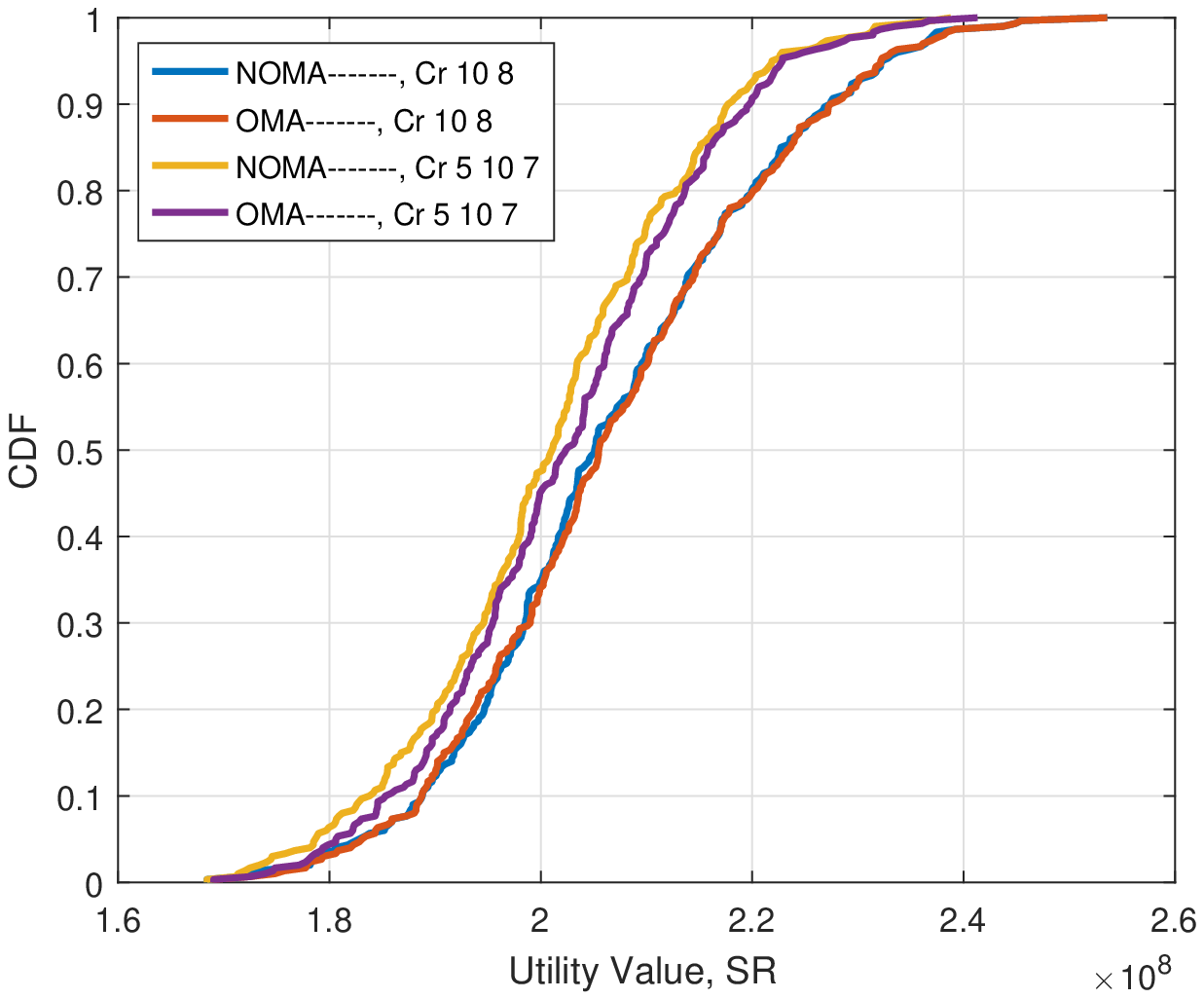}
         \caption{\small Sum-rate maximization.}
         \label{fig:OMA_sr}
 \end{subfigure}
    \caption{OMA vs. NOMA,  for different $\bar{C}_{f}$ values. }
    \label{fig:OMA_vs_NOMA}
\end{figure}

Fig.~\ref{fig:OMA_vs_NOMA} compares the proposed NOMA schemes with H-PA-PS algorithm to the conventional OMA scheme under different utility functions and for different values of the fronthaul capacity. For OMA scheme, every RB of each FAP is assigned to only one user served with full power. For the sake of fair comparison, users are optimally assigned to FAPs using the same algorithm (one user assigned to each RB) as for NOMA. Moreover, the power allocated to every RB is optimized using the same technique. The adopted assignment is Hungarian-based for both OMA and NOMA.

For weighted sum-rate maximization in Fig.~\ref{fig:OMA_wsr}, NOMA achieves better performance in the low utility region, indicating higher fairness for users with poorer channels, such as edge users, which is the most challenging case.
For sum-rate maximization in Fig.~\ref{fig:OMA_sr}, NOMA and OMA have comparable performance for high $\bar{C}_f$. For low $\bar{C}_f$, OMA is generally slightly better, confirming that the best strategy for maximizing sum-rate is giving full power to the strong user.


Table.~\ref{table:jain_NOMA_vs_OMA} compares the Jain's fairness index for NOMA and OMA, for weighted sum-rate and sum-rate and for different values of the fronthaul capacity.
 The table shows that the user fairness is the highest when proposed NOMA is adopted with weighted-sum rate maximization.

	\renewcommand{\arraystretch}{1.3}
	\begin{table}[!h]
\centering
\begin{tabular}{|c|c|c|c|c|}
\hline
Utility &  \multicolumn{2}{|c|}{Weighted sum-rate} & \multicolumn{2}{|c|}{Sum-rate}\\
\hline
$\bar{C}_{f}$    & $5 \times 10^{7}$  & $ 10^{8}$  &$5 \times 10^{7}$   & $ 10^{8}$\\
	\hline\hline
			NOMA		& 0.42& 0.42& 0.25& 0.22\\
					\hline
			OMA		& 0.31& 0.32& 0.24& 0.23\\
\hline
\end{tabular}
\caption{Jain's fairness indexes for weighted sum-rate and sum-rate maximization, NOMA vs. OMA.}
				\label{table:jain_NOMA_vs_OMA}
\end{table}

\subsection{Effect of number of RBs and Weight}
\begin{figure}[t!]
    \centering
    \begin{subfigure}[t]{0.48\textwidth}
\centering
    \psfrag{Hungaria Fairness}[c][c][0.6]{Jain's fairness index}
    \psfrag{r}[c][c][0.6]{Number of RBs per FAP, $r$}
    \psfrag{Cr---- 10 8}[c][c][0.47]{ $C_f=10^8$}
    \psfrag{Cr---- 5 10 7}[c][c][0.47]{ $C_f=5\times 10^7$}
    \psfrag{SR----}[c][c][0.47]{SR}
    \psfrag{WSR---- 25}[c][c][0.47]{WSR $\tau=$25}
    \psfrag{WSR---- 50}[c][c][0.47]{WSR $\tau=$50}
       \scalebox{1}{ \includegraphics[width=\textwidth]{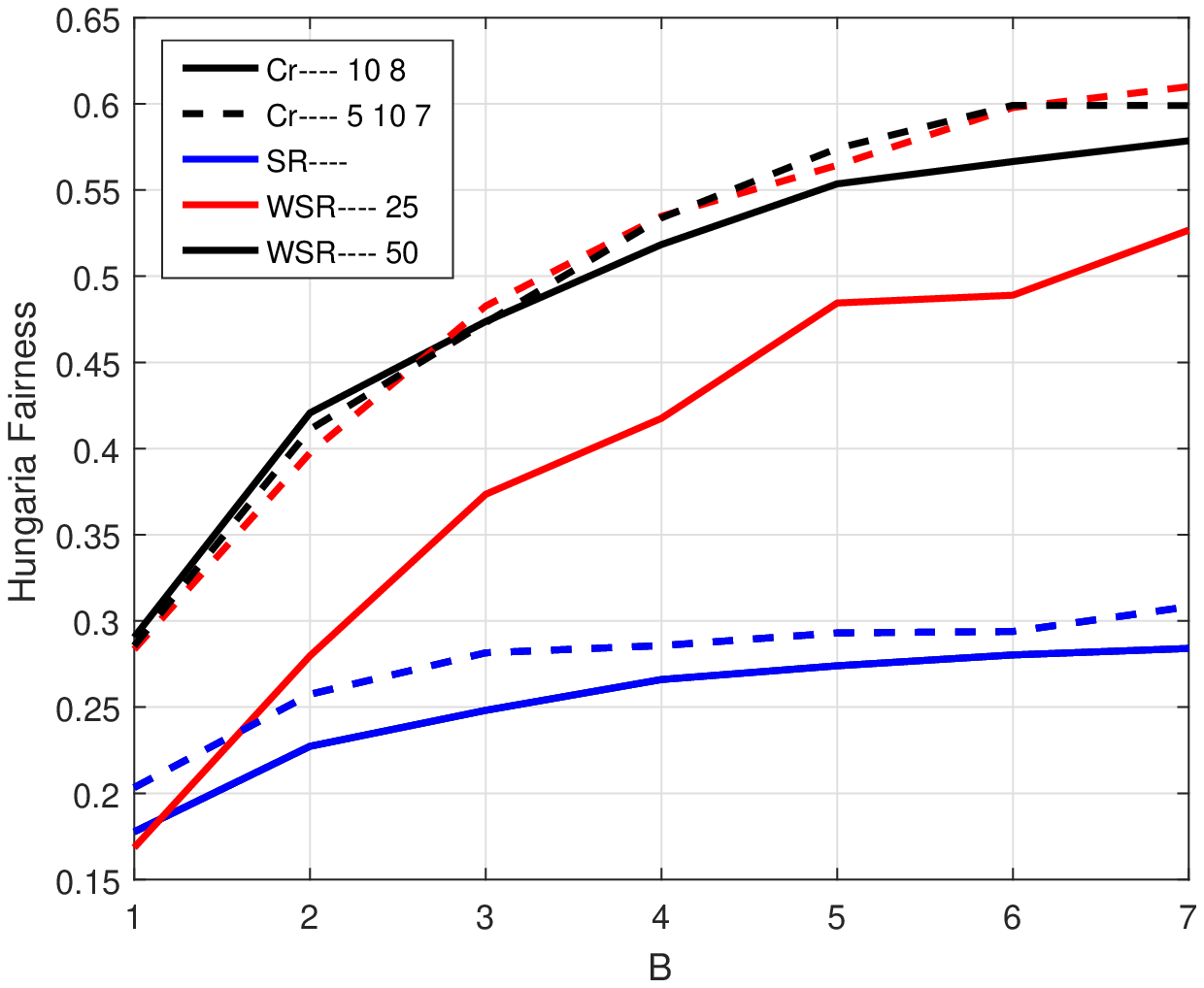}}
        \caption{Hungarian-based assignment.}
        \label{fig:fairnessRB_H}
    \end{subfigure}\hspace{-0.0\textwidth}
    ~
    \begin{subfigure}[t]{0.48\textwidth}
     \psfrag{MCKP Fairness}[c][c][0.6]{Jain's fairness index}
    \psfrag{r}[c][c][0.6]{Number of RBs per FAP, $r$}
    \psfrag{Cr---- 10 8}[c][c][0.47]{ $C_f=10^8$}
    \psfrag{Cr---- 5 10 7}[c][c][0.47]{ $C_f=5\times 10^7$}
    \psfrag{SR----}[c][c][0.47]{SR}
    \psfrag{WSR---- 25}[c][c][0.47]{$\quad$WSR $\tau=$25}
    \psfrag{WSR---- 50}[c][c][0.47]{$\quad$WSR $\tau=$50}
       \scalebox{1}{ \includegraphics[width=\textwidth]{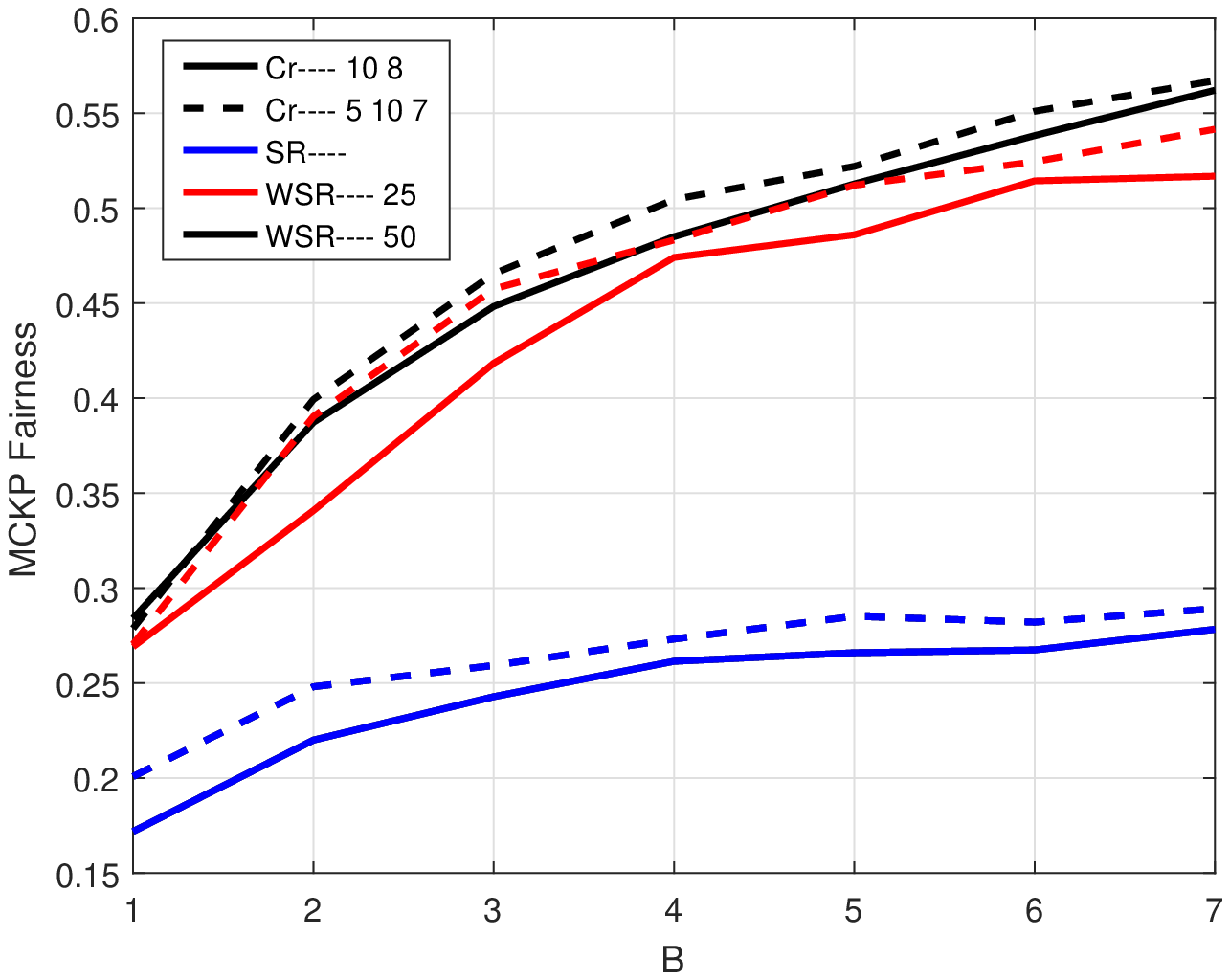}}
        \caption{MCKP-based assignment.}
        \label{fig:fairnessRB_K}
 \end{subfigure}
    \caption{Variation of the fairness with the number of RBs, for different utility functions. }
    \label{fig:fairnessRB}
\end{figure}

Fig.~\ref{fig:fairnessRB} shows the variation of the Jain's fairness index according to the number of RBs available at every FAP. The figure shows the fairness index for different values of the fronthaul capacity and for different utility functions, and with $\tau=25$, or 50. The results are presented for H-PA-PS in~\ref{fig:fairnessRB_H} and for K-PA-PS in~\ref{fig:fairnessRB_K}.

 In general, for both assignment algorithms under any value of $\bar{C}_f$ and any adopted utility function, the fairness is an increasing function of the number of RBs. More RBs implies not only more served users, but also a higher probability of each user being served, owing to channel diversity.
  However, a higher number of RBs entails higher complexity, as shown in Table~\ref{table:algoComplexity}: the complexity of every algorithm linearly increases with the number of RBs $R$.

Overall, the lower $\bar{C}_f$, the better the fairness, for the same reason as for Table \ref{table:jain_algorithm}. For $\tau=25$, in MCKP-based assignment algorithm (Fig.~\ref{fig:fairnessRB_K}), the fairness is higher compared to the one resulting from Hungarian-based algorithm, especially for a low number of RBs. This phenomenon is caused by the fact that, under MCKP, users are assigned to FAPs under fronthaul capacity constraints, which increases the probability that weak users get served.
It is also observed that, for weighted sum-rate, the larger the window size gets, the better level the fairness index reaches, which is compatible with the rationale behind adopting a proportional fairness approach.

\subsection{Discussion}
In this subsection, we compare the two assignment algorithms proposed in this paper.
On one hand, with Hungarian-based assignment, the users are assigned to FAPs and RBs regardless of the fronthaul capacity. Moreover, the result of the assignment is optimal for infinite fronthaul capacity. The transmission power for each RB is then adjusted to optimize again the utility function and to satisfy the fronthaul capacity. On the other hand, MCKP-based assignment allocates the resource to users in order to maximize the utility and to satisfy the fronthaul capacity. It can be seen in Fig.~\ref{fig:convergence} that, in terms of utility values, with the MCKP-based assignment, the overall algorithm performs better than with the Hungarian-based assignment for weighted sum-rate maximization. In terms of user fairness, however, the performance of both assignment algorithms are comparable (the Hungarian-based assignment performs slightly better), cf. Table~\ref{table:jain_algorithm}. For sum-rate maximization, the utility value is slightly better with the Hungarian-based assignment. Nevertheless, it is worth mentioning that MCKP-based assignment can be applied alone for uniform power without power optimization. In terms of computational complexity, it can be shown, based on Table.~\ref{table:algoComplexity}, that the Hungarian-based assignment is less complex than the MCKP-based assignment if 
\begin{equation}
F < \sqrt{\frac{AU\left(U-1\right)}{32}\log\left(\frac{RU\left(U-1\right)}{2}\right)}.
\end{equation}
Therefore, for a high number of users competing for the resource, it is advisable to use the Hungarian-based assignment.

\section{Conclusion}\label{sec:Conclusion}

This paper proposes sophisticated resource allocation algorithms in the DL of a NOMA-based FRAN with multiples RBs, in an effort to realize the high performance requirements  of next generation wireless systems. The paper considers maximizing a weighted sum-rate optimization problem in the realm of the studied NOMA-based FRAN. The optimization problem is solved by iterating between three steps: user-to-FAP-and-RB assignment (binary optimization), power allocation to RBs, and NOMA power split between weak and strong users within every RB. Two different user assignment methods are proposed to efficiently solve the binary part, namely the Hungarian-based and MCKP-based methods. The continuous part of the problem, on the other hand, is solved using an ADMM approach, which enables the FRAN-specific operation and reduces the overall algorithmic complexity. The simulation results show the effect of different levels of fronthaul capacity on the network performance. The results of the paper particularly show that, compared to conventional OMA, the proposed NOMA strategy under FRAN constraints increases user fairness without harming network sum-rate.
Future research directions on the topic would include optimizing a more generalized FRAN architecture empowered by device-to-device (D2D) communication, relaying, and caching at the network edge, so as to best meet the ambitious requirements of the anticipated futuristic systems.

\bibliographystyle{IEEEtran}
\bibliography{citation2}
					
\end{document}